\documentclass[twocolumn,times,tighten]{aastex63}

\usepackage{apjfonts}
\usepackage{graphicx}
\usepackage{xcolor}
\usepackage{amsmath}	
\usepackage{amssymb}	
\usepackage{gensymb}
\usepackage{bm}         
\hypersetup{breaklinks=true}
\usepackage{natbib}
\setcitestyle{aysep={}, citesep={;}}

\newcommand{\soft}[1]{{\scshape{#1}}}

\newcommand{\ie}{i.e.,}

\newcommand{\inv}{\ensuremath{^{-1}}}

\newcommand{\zsim}[1]{\ensuremath{z \simeq{} {#1}}}
\newcommand{\zeq}[1]{\ensuremath{z = {#1}}}
\newcommand{\LSun}{\ensuremath{L_{\odot{}}}}

\renewcommand{\vec}{\boldsymbol}

\thickmuskip=\medmuskip
\medmuskip=\thinmuskip

\makeatletter
\@ifclassloaded{emulateapj}{}{}
\@ifclassloaded{emulateapj}{}{}
\@ifclassloaded{emulateapj}{}{}
\makeatother

\usepackage[hang,flushmargin]{footmisc}

\begin{document}

\title{Limits to Rest-Frame Ultraviolet Emission From Far-Infrared-Luminous \boldmath\(z\simeq{}6\) Quasar Hosts}

\author[0000-0001-6434-7845]{M. A. Marshall}
\affiliation{School of Physics, University of Melbourne, Parkville, VIC 3010, Australia}
\affiliation{ARC Centre of Excellence for All Sky Astrophysics in 3 Dimensions (ASTRO 3D)}

\author[0000-0001-6462-6190]{M. Mechtley}
\affiliation{School of Earth and Space Exploration, Arizona State University, P.O. Box 871404, Tempe, AZ 85287, USA}

\author[0000-0001-8156-6281]{R. A. Windhorst}
\author[0000-0003-3329-1337]{S. H. Cohen}
\author[0000-0003-1268-5230]{R. A. Jansen}
\affiliation{School of Earth and Space Exploration, Arizona State University, P.O. Box 871404, Tempe, AZ 85287, USA}

\author[0000-0003-4176-6486]{L. Jiang} 
\affiliation{The Kavli Institute for Astronomy and Astrophysics, Peking University, Beijing, 100871, People's Republic of China}

\author[0000-0003-4665-8521]{V. R. Jones}
\affiliation{Steward Observatory, The University of Arizona, Tucson, AZ 85721, USA}

\author[0000-0001-7956-9758]{J. S. B. Wyithe}
\affiliation{School of Physics, University of Melbourne, Parkville, VIC 3010, Australia}
\affiliation{ARC Centre of Excellence for All Sky Astrophysics in 3 Dimensions (ASTRO 3D)}

\author[0000-0003-3310-0131]{X. Fan}
\affiliation{Steward Observatory, The University of Arizona, Tucson, AZ 85721, USA}

\author[0000-0001-6145-5090]{N. P. Hathi} 
\affiliation{Space Telescope Science Institute, Baltimore, MD 21218, USA}

\author[0000-0003-3804-2137]{K. Jahnke} 
\affiliation{Max-Planck-Institut f\"ur Astronomie, K\"onigstuhl 17, 69117 Heidelberg, Germany}

\author[0000-0002-6131-9539]{W. C. Keel} 
\affiliation{Department of Physics and Astronomy, University of Alabama, Tuscaloosa, AL 35487, USA}

\author[0000-0002-6610-2048]{A. M. Koekemoer}
\affiliation{Space Telescope Science Institute, Baltimore, MD 21218, USA}

\author[0000-0003-1733-9281]{V. Marian}
\affiliation{Max-Planck-Institut f\"ur Astronomie, K\"onigstuhl 17, 69117 Heidelberg, Germany}
\affiliation{International Max Planck Research School for Astronomy \& Cosmic Physics at the University of Heidelberg, Germany}

\author[0000-0001-7347-5953]{K. Ren}
\affiliation{School of Physics, University of Melbourne, Parkville, VIC 3010, Australia}
\affiliation{ARC Centre of Excellence for All Sky Astrophysics in 3 Dimensions (ASTRO 3D)}

\author{J. Robinson}
\affiliation{School of Earth and Space Exploration, Arizona State University, P.O. Box 871404, Tempe, AZ 85287, USA}

\author[0000-0001-8887-2257]{H. J. A. R{\"o}ttgering}
\affiliation{Leiden Observatory, Postbus 9513, NL-2300 RA Leiden, The Netherlands}

\author[0000-0003-0894-1588]{R. E. Ryan, Jr.}
\affiliation{Space Telescope Science Institute, Baltimore, MD 21218, USA}

\author[0000-0002-3193-1196]{E. Scannapieco}
\affiliation{School of Earth and Space Exploration, Arizona State University, P.O. Box 871404, Tempe, AZ 85287, USA}

\author{D. P. Schneider}
\affiliation{Department of Astronomy and Astrophysics, The Pennsylvania State University, University Park, PA 16802, USA}
\affiliation{Institute for Gravitation and the Cosmos, The Pennsylvania State University, University Park, PA 16802, USA}

\author[0000-0002-4511-5966]{G. Schneider}
\affiliation{Steward Observatory, The University of Arizona, Tucson, AZ 85721, USA}

\author[0000-0002-0648-1699]{B. M. Smith}
\affiliation{School of Earth and Space Exploration, Arizona State University, P.O. Box 871404, Tempe, AZ 85287, USA}

\author[0000-0001-7592-7714]{H. Yan}
\affiliation{Department of Physics and Astronomy, University of Missouri, Columbia, MO, USA}

\begin{abstract}
We report on a Hubble Space Telescope search for rest-frame ultraviolet
emission from the host galaxies of five far-infrared-luminous $z\simeq{}6$
quasars and the $z=5.85$ hot-dust free quasar SDSS J0005-0006. 
We perform 2D surface brightness modeling for each quasar using a Markov-Chain Monte-Carlo estimator, to simultaneously fit and subtract the quasar point source in order to constrain the underlying host galaxy emission. We measure upper limits for the quasar host galaxies of $m_J>22.7$ mag and $m_H>22.4$ mag, corresponding to stellar masses of $M_\ast<2\times10^{11}M_\odot$. These stellar mass limits are consistent with the local $M_{\textrm{BH}}-M_\ast$ relation. 
Our flux limits are consistent with those predicted for the UV stellar populations of $z\simeq6$ host galaxies, but likely in the presence of significant dust ($\langle A_{\mathrm{UV}}\rangle\simeq 2.6$ mag).
We also detect a total of up to 9 potential $z\simeq6$ quasar companion galaxies surrounding five of the six quasars, separated from the quasars by $1\farcs4$--$3\farcs2$, or 8.4--19.4 kpc, which may be interacting with the quasar hosts. These nearby companion galaxies have UV absolute magnitudes of $-22.1$ to $-19.9$ mag, and UV spectral slopes $\beta$ of $-2.0$ to $-0.2$, consistent with luminous star-forming galaxies at $z\simeq6$. 
These results suggest that the quasars are in dense environments typical of luminous $z\simeq6$ galaxies.
However, we cannot rule out the possibility that some of these companions are foreground interlopers. Infrared observations with the James Webb Space Telescope will be needed to detect the $z\simeq6$ quasar host galaxies and better constrain their stellar mass and dust content.
\end{abstract}


\section{Introduction}\label{introduction}

Since their initial discovery in the Sloan Digital Sky Survey
\citep[SDSS,][]{fan_2000, fan_2001, fan_2003, fan_2004},
high-redshift ($z\gtrsim6$) quasars have been invaluable probes of the early Universe.
These quasars can constrain black hole seed theories
\citep{mortlock_2011,Volonteri2012,Banados2017}, the reionization history of the Universe
\citep{Fan2006a,mortlock_2011,Greig2016,Davies2018,Greig2019}, and provide unique insights
into the connection between black hole and galaxy growth at the end of the Epoch
of Reionization \citep{Shields2006,Wang2013,Valiante2014,Schulze2014,Willott2017}.

The extreme nature of these objects, with large black hole masses
\citep[$M_{\rm BH}\simeq10^9M_\odot$;][]{Barth2003,Jiang2007,Kurk2007,DeRosa2011}
and accretion rates near and even above the Eddington limit
\citep{willott_2010,DeRosa2011}, suggests that these quasars may live in extreme
high-density environments. 
However, observations do not find that quasars reside in high-density regions
\citep[e.g.][]{Kim2009,Banados2013,Morselli2014}, challenging our understanding.
Theoretically, however, it is kpc-scale interactions that could trigger supermassive black
hole growth
\citep[e.g.,][]{Sanders1988,Hopkins2006}, despite not universally being observed in
lower redshift ($z<2$) quasar systems
\citep{Cisternas2011,Kocevski2012,Mechtley2016,Villforth2018,Marian2019}. Recently, Atacama Large Millimeter/submillimeter Array (ALMA) observations
in the sub-mm have detected galaxies around high-redshift quasars at separations of
$\sim8$--60 kpc \citep{Decarli2017,Trakhtenbrot2017}, which have been interpreted as
major galaxy interactions. These observations suggest that major mergers may be important drivers
of rapid black hole
growth in the early Universe, and thus observations
must probe the local environments of quasars to understand these extreme systems.

Alongside their local environment, many studies investigate the host galaxies of these
quasars to understand the connection between black hole and galaxy growth in the early
Universe \citep[e.g.][]{Shields2006,Wang2013,Willott2017}. However, observations of
quasar host galaxies are challenging with current facilities \citep[e.g.][]{Bahcall_1994,disney_1995,kukula_2001,hutchings_2003}.
These observations are strongly focused in two wavelength ranges where detectability
is relatively easy: rest-frame ultraviolet (UV) emission observed in the near-infrared from
$\approx{}0.7-2.2\mu{}$m, and rest-frame far-infrared (FIR) emission observed at sub-mm wavelengths. The UV emission traces the bright
accretion disk and stellar light from the host galaxy, while the FIR instead
predominantly traces cold dust in the host. 

The extreme luminosity of quasars
in the UV often means that they significantly outshine their hosts \citep[e.g.,][]{schmidt_1963,mcleod_1994,dunlop_2003,hutchings_2003,floyd_2013}. 
The highest redshift at which the UV emission from a quasar host has unambiguously been
observed from ground-based telescopes is $z\simeq4$ \citep{mcleod_2009,targett_2012}.
Galaxies are more compact at higher redshifts, with physical sizes evolving as $R_e\propto (1+z)^{-m}$, where $m$ is typically measured to be between 1 and 1.5 \citep[e.g.][]{Bouwens2004,oesch_2010,Ono2013,Kawamata2015,Shibuya2015,Laporte2016,Kawamata2018}.
This rate of decrease of galaxy sizes toward higher redshifts is stronger than the increase in apparent diameters at $z$ $\gtrsim$ 2 due to the cosmic angular size--distance relation.
Thus, at higher redshifts, the angular size of galaxies becomes small relative to the point spread function (PSF) of current telescopes, and so the bright quasar entirely conceals the host galaxy emission \citep[e.g.][]{Mechtley2012}. 
Surface brightness dimming also causes the host galaxies and any tidal features to be more difficult to detect at high redshift.

In an attempt to detect the underlying UV emission from the host of the redshift $z = 6.42$ quasar SDSS J114816.64+525150.3
(hereafter SDSS J1148+5251), \citet{Mechtley2012} used
GALFIT \citep{peng_2010} to model the quasar contribution to the emission in Hubble Space
Telescope (HST) Wide Field Camera 3 (WFC3) images. This quasar model was subtracted to obtain upper limits on the
brightness of the host galaxy, of $m_J>22.8$ and $m_H>23.0$ mag. To improve the fitting method, \citet{Mechtley2014}
developed a Markov-Chain Monte-Carlo (MCMC) simultaneous fitting software,
\soft{psfMC}\footnote{The details of the software implementation are given in
\citet{Mechtley2014}. The software, documentation, examples, and source code are
available at: \url{https://github.com/mmechtley/psfMC}}.
While this technique allows for host detections at lower redshifts \citep[$z=2$,][]{Mechtley2016,Marian2019}, the smaller angular sizes of the hosts at
higher redshifts make this significantly more challenging. 

In this paper, we present deep near-infrared F125W (J) and F160W (H) HST WFC3 images of six $z\simeq6$ quasars. We
describe our efforts to detect rest-frame near-UV
emission from the hosts, and present the most robust upper limits
to date on the rest-frame UV brightness of each of the quasar host galaxies. This significantly increases the sample of high-redshift quasar hosts with deep UV upper limits determined by this method, extending on the previous work of \citet{Mechtley2012} which studied only one quasar. 
The subtraction of the quasar PSF using
the \soft{psfMC} software also allows for an unobscured view of the quasar
environment on kpc-scales, uncovering nearby galaxies which may be interacting with the
host and triggering this rapid black hole growth.


Throughout this paper we adopt a \(\Lambda{}\)CDM cosmology with
\(H_{0}=67\)~km~s\inv{}~Mpc\inv{}, \(\Omega{}_{M} = 0.3\), and
\(\Omega{}_{\Lambda{}} = 0.7\) \citep{Planck2014}. All magnitudes are on the AB system \citep{oke_1983} and have been corrected for
Galactic extinction using the reddening map of
\citet{schlegel_1998} as recalibrated by \citet{Schlafly2011}.

\section{Quasar Sample}\label{quasar-sample}

In this work we study five UV-faint FIR-luminous quasars and one dust-free quasar,
all at $z\simeq6$. The dust-free quasar was observed in a second epoch of the original
pilot program, alongside SDSS J1148+5251 \citep[ID 12332, PI: R.~Windhorst;
see][]{Mechtley2012}, but is previously unpublished. The five UV-faint FIR-luminous
quasars were observed in 2013 as part of HST program 12974
(PI: M.~Mechtley), which built on the original program. The observations and modeling technique
(\S{}~\ref{hubble-space-telescope-data-and-observing-strategy}--\ref{source-modeling-and-point-source-subtraction})
are identical for all sources. Relevant properties of
each of the six targets are summarized in Table~\ref{tab_targets}.

\subsection{UV-faint FIR-luminous Quasars}\label{uv-faint-fir-luminous-quasars}

Guided by our initial experience with
\mbox{SDSS J1148+5251} \citep{Mechtley2012}, we determined that high-redshift quasars with weaker UV emission ($M_{1450\textrm{\AA}}>-26.5$ mag), but secure
sub-mm detections, i.e., with large rest-frame FIR to UV flux
ratios ($F_{\rm FIR}/F_{\rm UV}\gtrsim100$), are the best candidates for successful
detection of host emission. 

\begin{figure}
\centering
\includegraphics[width=\columnwidth]{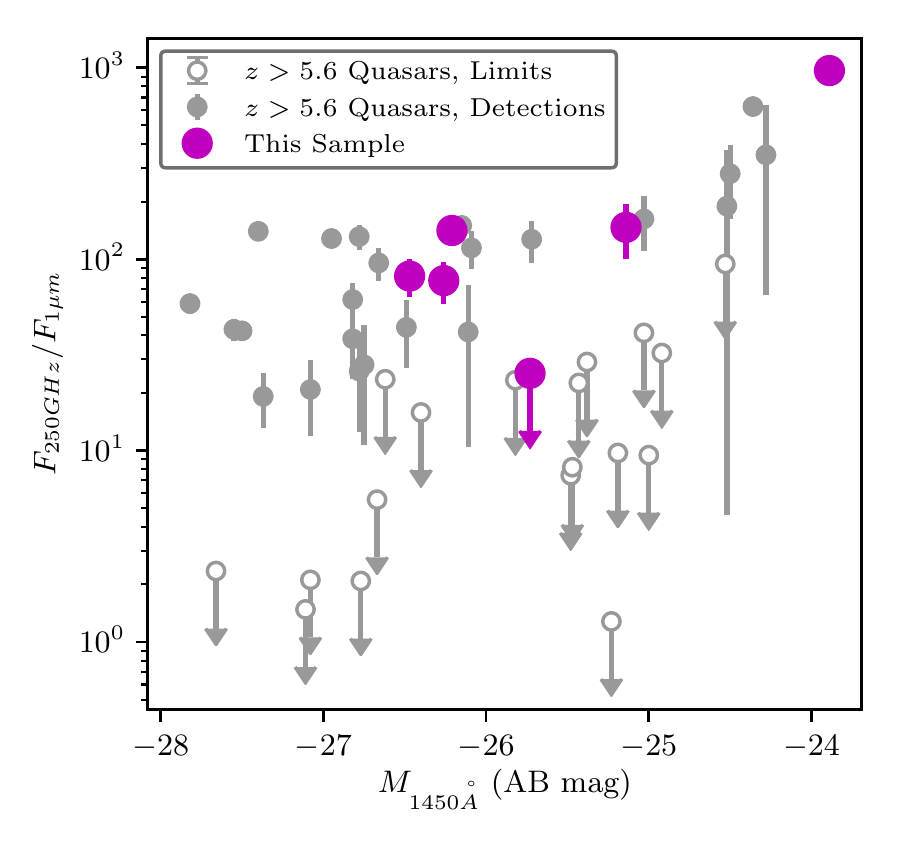}
\caption{Selection of ultraviolet-faint, far infrared-luminous quasars based on
absolute magnitude (rest-frame 1450\AA{}) and observed sub-mm to near-infrared
flux ratio. Our sample of six quasars is denoted by magenta circles, with detections for the five IR-luminous quasars, and an upper limit for the additional quasar SDSS J0005-0006. Other $z>5.6$ quasars with sub-mm observations are plotted in gray
\citep{fan_2000,fan_2001,fan_2003,fan_2004,Fan2006,petric_2003,bertoldi_2003,mahabal_2005,Cool2006,mcgreer_2006,goto_2006,venemans_2007,venemans_2013,Wang2007,Wang2008,Wang2011,Wang2013,Kurk2007,kurk_2009,willott_2007,willott_2010,willott_2010b,jiang_2007,jiang_2008,jiang_2009,mortlock_2009,mortlock_2011,zeimann_2011,DeRosa2011,de_rosa_2014,omont_2013,Banados2014,wu_2015}.
}
\label{fig:submm_selection}
\end{figure}

The rationale behind this selection is that a high FIR luminosity---and associated high star formation rate---coupled with a lower
nuclear UV luminosity results in a less extreme
nuclear-to-host contrast ratio, and thus improved detectability of host UV emission. At the
time of selection (February 2012), there were only five such
\zsim{6} quasars known that met these criteria:
\mbox{CFHQS J0033$-$0125}, \mbox{SDSS J0129$-$0035}, \mbox{SDSS J0203+0012},
\mbox{NDWFS J1425+3254}, and {SDSS} \mbox{J2054} \mbox{$-$0005} (see Figure~\ref{fig:submm_selection}).
We note that, while these quasars are UV-`faint' relative to the observed high-redshift quasar sample, they are still very luminous in the UV with $-26.5<M_{1450\textrm{\AA}}<-23.9$ mag.

Although the FIR emission suggests the presence of significant dust in
the host galaxies, the quasar discovery spectra (rest-frame UV) do not
show anomalous features compared to the rest of the population---\ie{}
they are otherwise normal \zsim{6} quasars, rather than showing
significant spectral reddening or absorption features such as present in the FIRST/2MASS sample at lower redshifts \citep{Urrutia2008,Glikman2015}. Furthermore, more than
$\sim25\%$ of \zsim{6} quasars have similarly high FIR
luminosities \citep{willott_2007, Wang2008, Wang2010, Wang2011},
so these FIR-luminous quasars are broadly representative of a
significant sub-population, rather than atypical objects.

\subsection{Dust-Free Quasar}\label{dust-free-quasar}

In addition to the FIR-luminous quasars described above, we also analyze
data from the prototype hot-dust-free quasar \mbox{SDSS J0005--0006}
\citep{fan_2004, jiang_2010}, which also lacks cold dust \citep[][]{Wang2008}. With a lower-luminosity and no evidence for
significant dust content, this quasar was selected as a counterpoint to
\mbox{SDSS J1148+5251}. 
This source is representative of a smaller, but still important
sub-population. At $5.8<z<6.4$, \citet{jiang_2010} found two
apparently dust-free quasars in a sample of 21 quasars, or $\approx 10\%$ of the population. \cite{leipski_2014} also found that $\approx 15\%$ of their sample of 69 quasars at $z>5$ are deficient in (but not devoid of) hot dust, and there is evidence of a trend toward higher dust-poor fraction with increasing redshift \citep{Jun2013}.

\begin{deluxetable*}{lllrr}
\tablecolumns{5}
\tablewidth{0pt}
\tablecaption{Quasars Observed with HST}
\tablehead{
\colhead{Quasar Name} &
\colhead{Redshift} &
\colhead{$M_{1450}$ (mag)} &
\colhead{$L_{FIR}$ $(10^{12}~\LSun{})$} &
\colhead{$\log(M_{BH}/M_\odot)$}
}
\startdata
CFHQS J003311.40--012524.9 & 6.13   & $-25.14$ & $2.6\pm{}0.8$ & $9.52\pm{}0.87^a$ \\
SDSS J012958.51--003539.7  & 5.78 & $-23.89$ & $5.2\pm{}0.9$ & $8.23\pm{}0.45^b$ \\
SDSS J020332.39+001229.3  & 5.72   & $-26.26$ & $4.4\pm{}1.1$ & $10.72\pm{}0.26^a$\\
NDWFS J142516.30+325409.0 & 5.89 & $-26.47$ & $5.4\pm{}1.2$ & $9.41\pm{}0.11^a$ \\
SDSS J205406.42--000514.8  & 6.04 & $-26.21$ & $5.5\pm{}1.2$ & $8.95\pm{}0.47^b$ \\
SDSS J000552.34--000655.8  & 5.85   & $-25.73$ & $<3.4$ & $8.02^c$ \\
\enddata
\tablecomments{Quasar names include the full sexagesimal coordinates. Redshifts
and absolute magnitudes use the same references as Table 7 in
\citet{Banados2016}. FIR luminosities are from \citet{Wang2010, Wang2011}. Black hole masses are from $a$) \citet{Shen2019}, $b$) \citet{Wang2013}/\citet{Willott2015} and $c$) \citet{Trakhtenbrot2017a}, and are calculated using the MgII line where available, else with the CIV line (NDWFS-J1425+3254) or by assuming the black hole is accreting at the Eddington luminosity (SDSS J0129-0035 and SDSS J2054-0005).}
\label{tab_targets}
\end{deluxetable*}

\section{Hubble Space Telescope Data and Observing
Strategy}\label{hubble-space-telescope-data-and-observing-strategy}

Each of the six quasars was observed with the HST WFC3 infrared channel in the F125W (J-band) and F160W (H-band) filters. 
The five FIR-luminous quasars were observed for two orbits (4800\,s) in each filter, while \mbox{SDSS J0005--0006} was observed for four orbits (10400\,s) in each filter. 
\citet{Windhorst2011} provides details on the WFC3 IR two-orbit sensitivity.

In addition to the quasar observations, coeval observations of a nearby PSF reference star were completed along with each epoch of quasar
imaging. Although the HST PSF is stable compared to ground-based
observatories, slight changes in the position of the secondary mirror
cause small time-dependent focus variations. These variations are believed to be
caused primarily by changes in the spacecraft thermal environment
\citep{bely_1993, hershey_1998, cox_2011}.
We mitigated this effect by imposing constraints on the PSF star
observations, as in the pilot program
\citep{Mechtley2012}---the (non-binary) stars were selected to be
within \(5\degr{}\) of the quasar, to minimize differences in the solar
illumination angle,
and the stars were observed in the orbit immediately
following the quasar observations, to best match the orbital day/night
cycle. 
The HST flight calendar builders also attempted, where
possible, to schedule our quasar and PSF observations {\it immediately} after an HST
target from a different program in a similar part of the sky as our quasar, so as to
further mitigate differences in orbital thermal variations between our first and
subsequent orbits on that quasar and PSF target. This special request was possible to
schedule for some of our quasars.
PSF star exposures were alternated in F125W and F160W to fully
sample the focal variation within an orbit \citep[for details, see][]{Mechtley2012}. Additionally, the stars were
selected to have (\(J-H\)) colors similar to the quasars, since the
diffraction-limited PSF also varies with wavelength. In wide filters,
redder sources can have a measurably broader PSF than bluer sources.

Four exposures were taken in each orbit of quasar and PSF star observations,
using the four-point box sub-pixel dither pattern to improve PSF
sampling and assist in the rejection of bad pixels and cosmic rays.
Critically-sampled images were reconstructed using \soft{astrodrizzle}, following approaches similar to those described in \citet{Koekemoer2002, koekemoer_2011, koekemoer_2013}, with a linear
pixel scale of $0\farcs{}06$ (a spatial scale of $\approx 0.36$ kpc at $z\simeq6$) and a \texttt{pixfrac} parameter of 0.8,
to reduce correlated noise while maintaining a relatively uniform weighting per-pixel. 
We used ``ERR'' (inverse variance) weighting for the final image combination step.
We transformed the ERR extensions from the HST exposures to per-pixel RMS error maps that include all sources of error, including shot noise, and account for correlated noise, as in \citet{Casertano2000} and \citet{Dickinson2004}, using \soft{astroRMS}\footnote{https://github.com/mmechtley/astroRMS}.

\section{Source Modeling and Point Source
Subtraction}\label{source-modeling-and-point-source-subtraction}

\begin{figure*}
\centering
\includegraphics[width=0.94\textwidth]{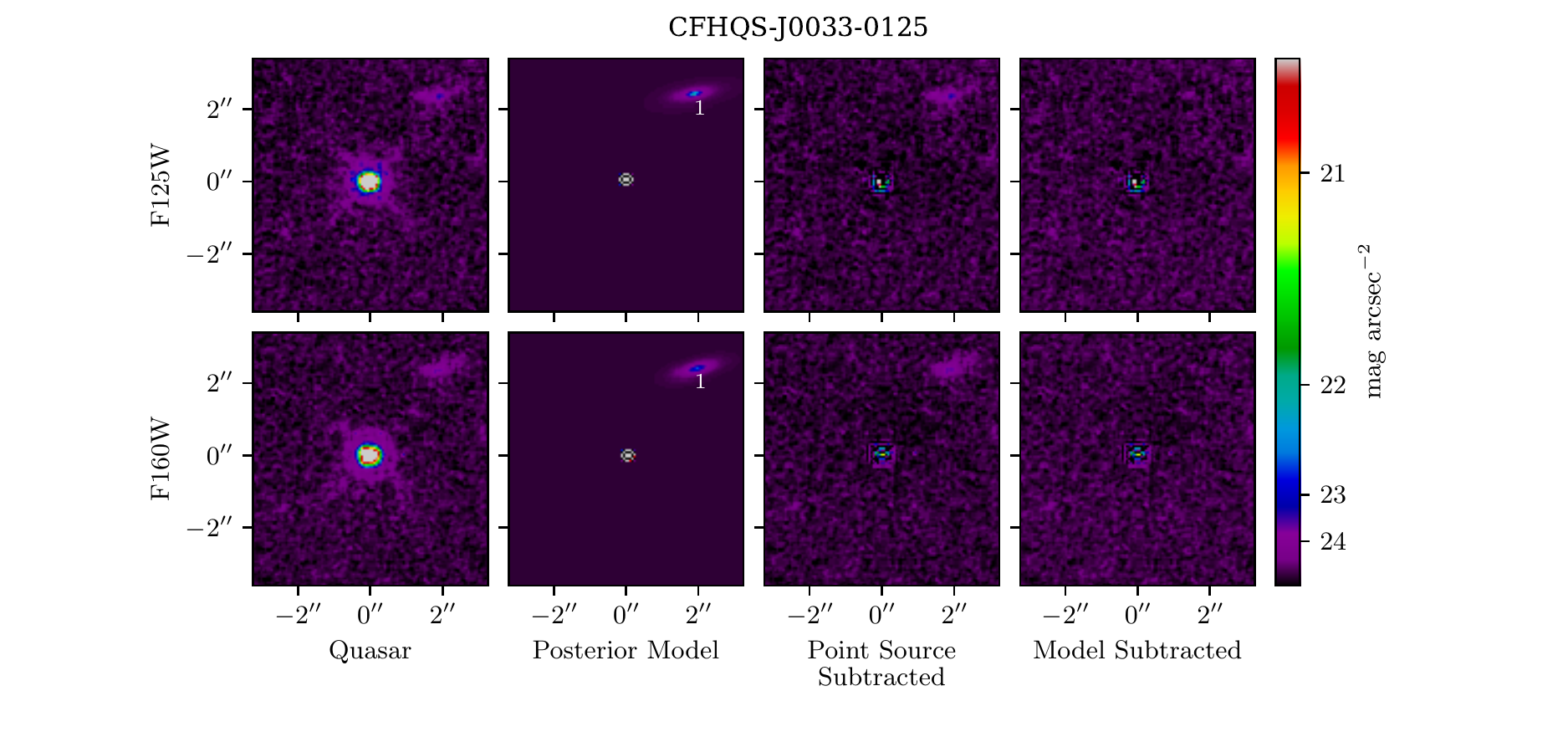}
\caption{Posterior-weighted model images for \mbox{CFHQS J0033--0125}. All images show a $\approx 6\farcs5 \times 6\farcs5$ FOV around the quasar, in order to see any companion galaxies, and are displayed with the same arcsinh color stretch and $0\farcs060$ pixel scale.
Top row: F125W filter. Bottom row: F160W filter. 
First column: drizzled, undistorted WFC3 images. 
Second column: posterior-weighted models from the MCMC
fitting process, before convolution with the PSF. 
Third column: residual after subtracting only the point source model from the original image. 
Fourth column: residual after subtracting the point source model and all modeled companions from the original image. Any galaxies surrounding the quasar are indicated with a white number, for ease of comparison with Figs. \ref{fig:colour_mag} and \ref{fig:size_mag} and Table \ref{tab:galaxy_properties}.
}
\label{fig:quasar1_resids}
\end{figure*}

\begin{figure*}
\centering
\includegraphics[width=0.94\textwidth]{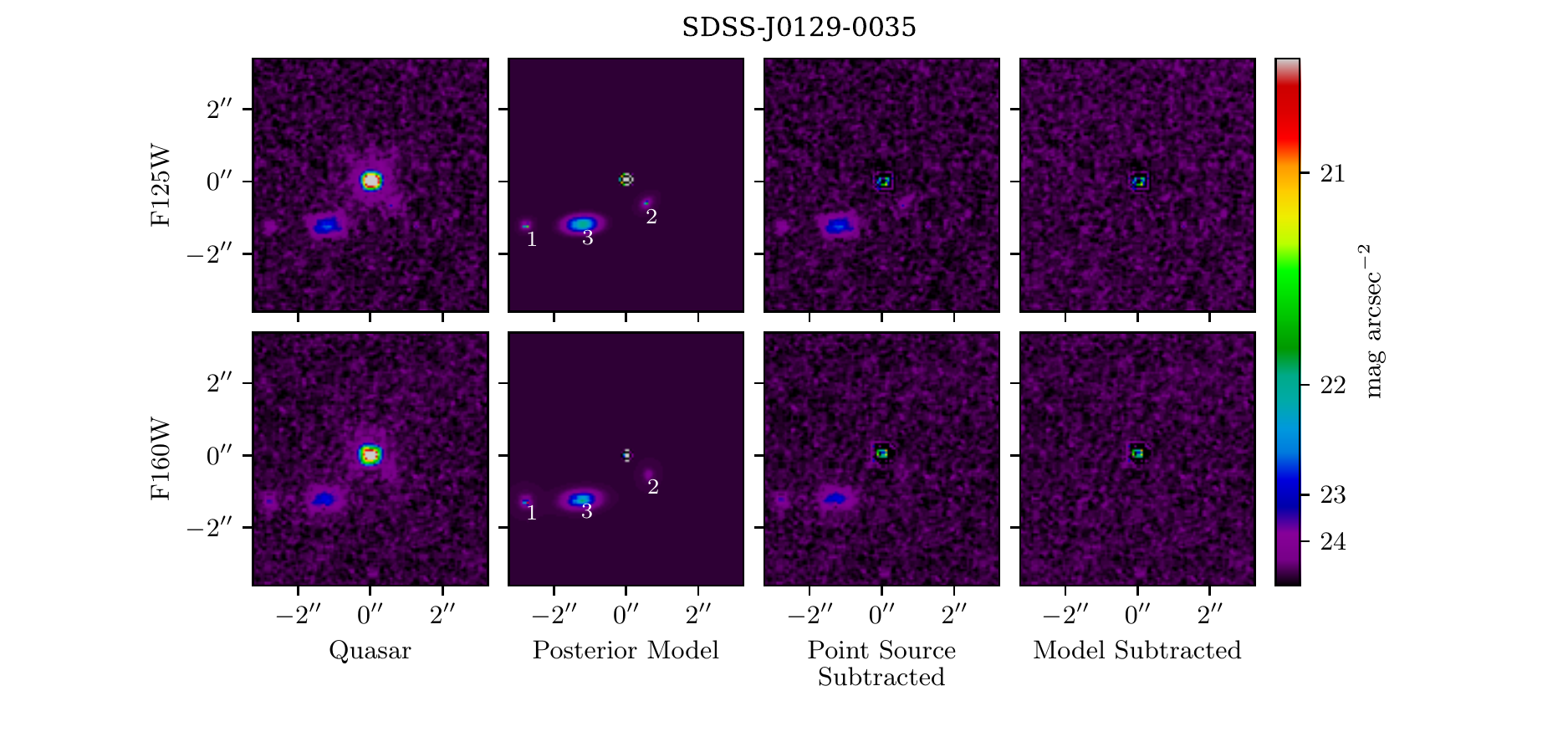}
\includegraphics[width=0.94\textwidth]{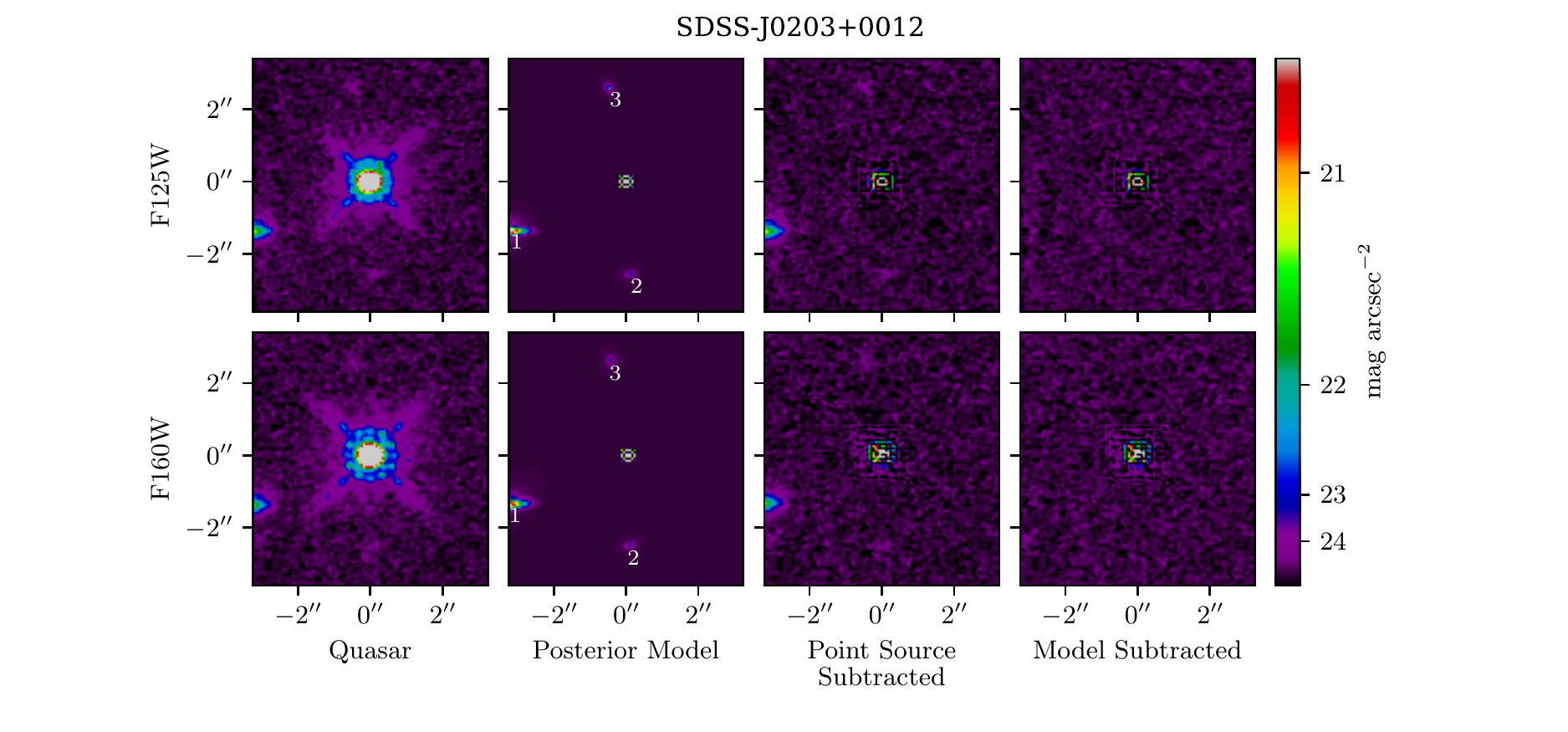}
\includegraphics[width=0.94\textwidth]{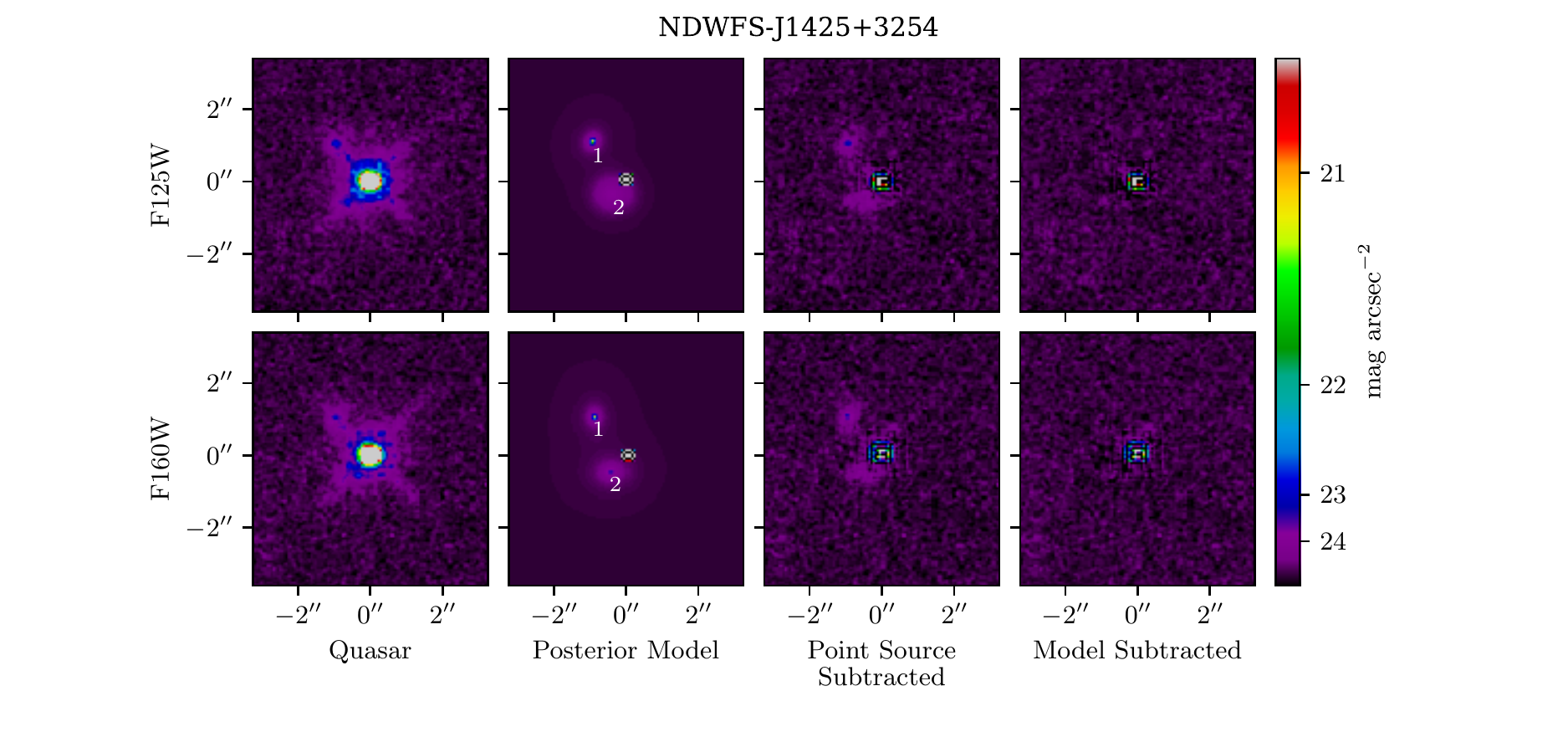}
\caption{Posterior-weighted model images for \mbox{SDSS J0129--0035}, \mbox{SDSS J0203+0012} and \mbox{NDWFS J1425+3254}. See
Figure~\ref{fig:quasar1_resids} for details.
}
\label{fig:quasar2_resids}
\end{figure*}

\begin{figure*}
\centering
\includegraphics[width=0.94\textwidth]{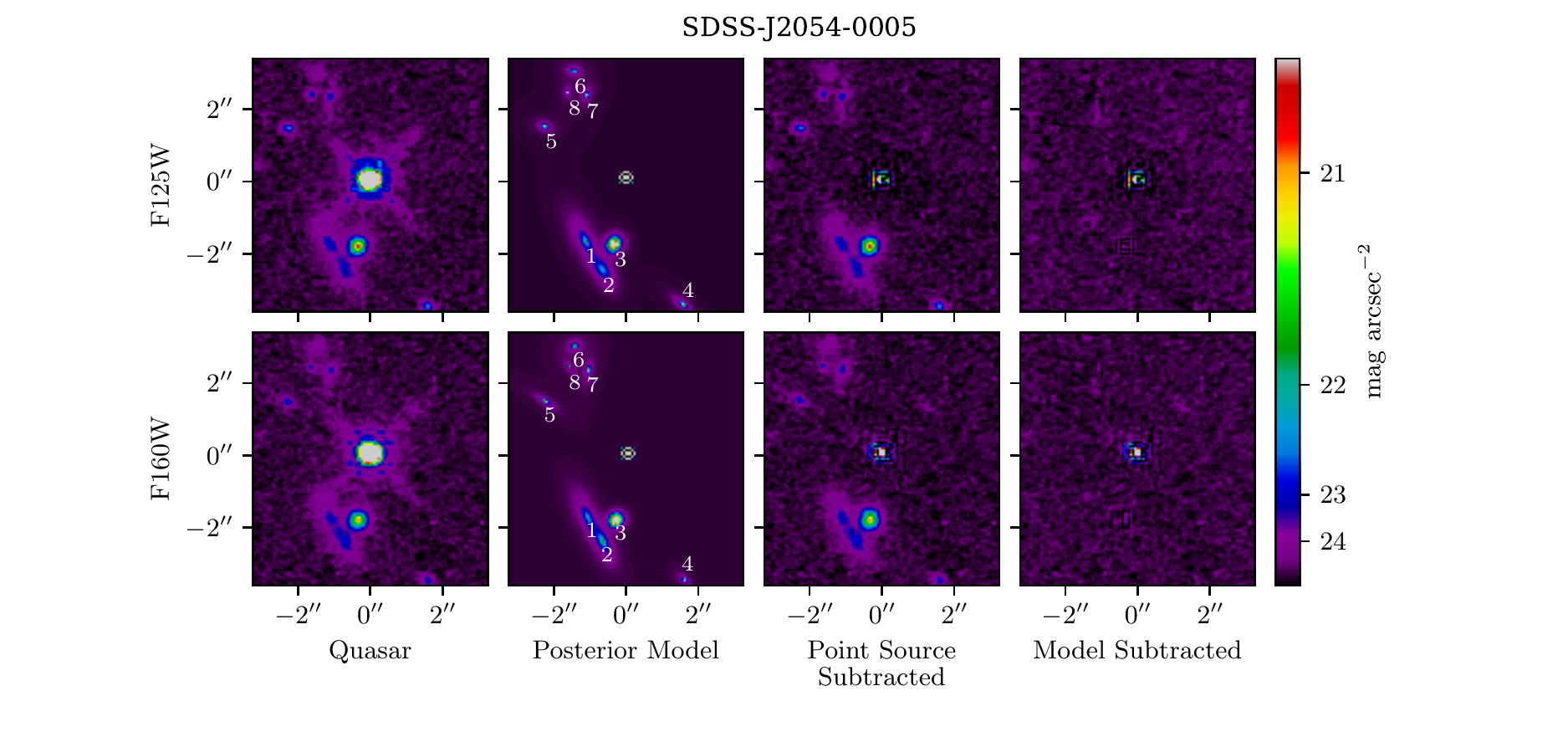}
\includegraphics[width=0.94\textwidth]{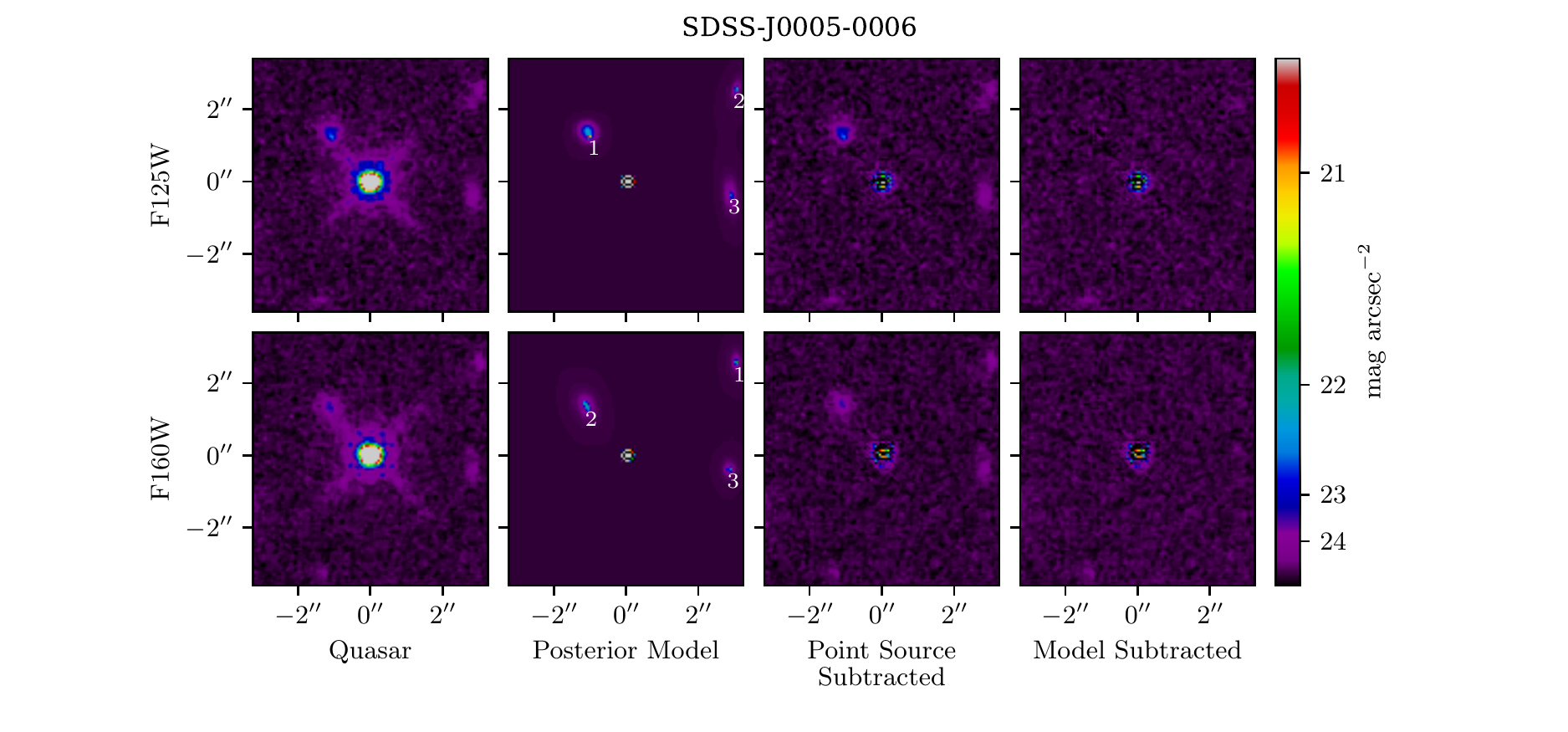}
\caption{Posterior-weighted model images for \mbox{SDSS J2054--0005} and \mbox{SDSS J0005--0006}. 
See
Figure~\ref{fig:quasar1_resids} for details.
}
\label{fig:quasar3_resids}
\end{figure*}

We performed 2D surface brightness modeling for each quasar using the
publicly available MCMC-based software \soft{psfMC}
\citep{Mechtley2014,Mechtley2016}. 
\soft{psfMC} allows the user to model an input image using a combination of point sources and \mbox{S\'ersic} profiles \citep{Sersic1963,Sersic1968} with the parameters: sky background; point source magnitude and position; and \mbox{S\'ersic} magnitude, position, \mbox{S\'ersic} index $n$, effective radius of the major axis $R_e$, ratio between the major and minor axes $b/a$, and position angle.
The MCMC process explores a range of model parameters specified by input prior probability distributions, convolving each model with an input PSF and comparing it with the telescope image, to determine the  posterior probability distribution of model parameters given the observed data.
The software uses the \soft{emcee} ensemble sampler \citep{emcee2013},
which improves sampling efficiency compared to the \soft{pyMC} \citep{patil_2010} version that was used in \citet{Mechtley2014}.

For each image of each source, we attempted
two different models---one with both a point source and an underlying
\mbox{S\'ersic} profile, and one with only a point source. We compared
the results of the two models both visually and using the Bayesian
Information Criterion as a model selection heuristic. \textit{In all cases},
there was no evidence that the data required the additional
\mbox{S\'ersic} profile---the seven additional free parameters were
primarily fitting noise peaks rather than residual flux from the hosts.

For all further analysis, we model the quasar as a pure point source. We also model any surrounding galaxies within $\approx3''$ of the quasar with a \mbox{S\'ersic} profile\footnote{Two galaxies could not be reasonably fit by one \mbox{S\'ersic} profile, so we instead fit them with two \mbox{S\'ersic} profiles superimposed, constraining their \mbox{S\'ersic} indices such that one represents a disk-like component and the other a more spheroidal component. The properties of both profiles for these galaxies are given in Table \ref{tab:galaxy_properties}, with their UV magnitude and slope calculated using the combined magnitude of both profiles.}. It should be stressed, however, that if the galaxies are associated with the quasar and undergoing a merger, their rest-frame UV emission need not be distributed in anything like a S\'ersic profile. Rather, this approach is simply used to model their flux to avoid over-subtraction. We assume uniform priors over a reasonable range, for all of the model parameters.
For each quasar image, we run the MCMC with 200 chains, and a minimum of 10,000 iterations with the first 5,000 discarded as a burn-in period (systems with more surrounding galaxies required up to twice as many iterations to obtain convergence). 
To ensure that the model is well-fit to the data, we examine the resulting posterior distributions, altering the allowed parameter range and iteration count until each parameter has converged and the residual flux in the model subtracted image is consistent with random noise.

For the six quasars and their companion galaxies, we create posterior-weighted model images before convolution with the PSF, and after the model has been convolved with the PSF and subtracted from the original image.
These weighted images are the (per-pixel) mean of all sample images, with more probable locations in parameter space being more densely populated with samples.
The resulting images for the six quasars in the J- and H-bands are shown in Figures \ref{fig:quasar1_resids}--\ref{fig:quasar3_resids}. The residual images show a central core of flux, which contains some residual flux from the quasar, and may also contain underlying host emission. The regions in the residual images where companion galaxy models have been subtracted purely consists of noise, demonstrating that the \soft{psfMC} model fits the observations superbly and our observations are noise-limited.

\soft{psfMC} also outputs the `best' parameter values from the maximum posterior model, alongside their errors. These values for the companion galaxy fits are given in Table \ref{tab:galaxy_properties}.

\section{Results}

\subsection{Quasar Host Galaxies}
\subsubsection{Magnitude Limits}

\begin{figure*}
\centering
\includegraphics[scale=0.8]{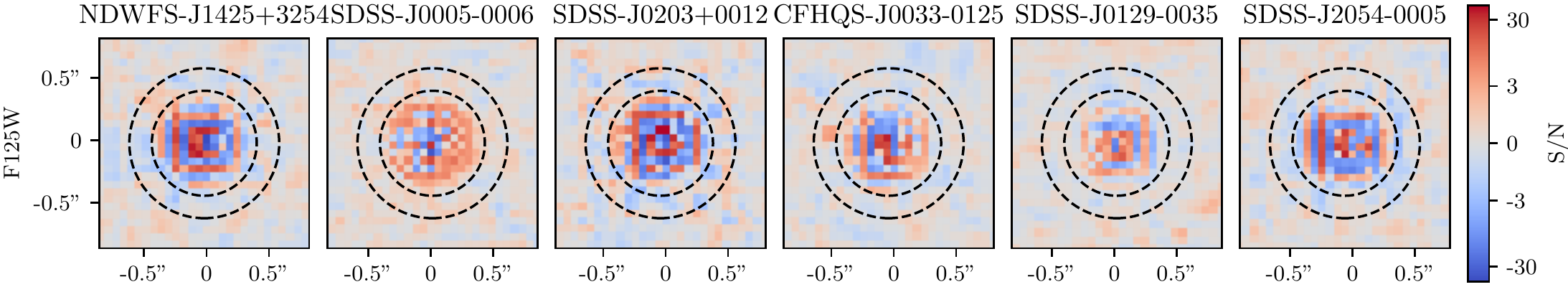}
\includegraphics[scale=0.8]{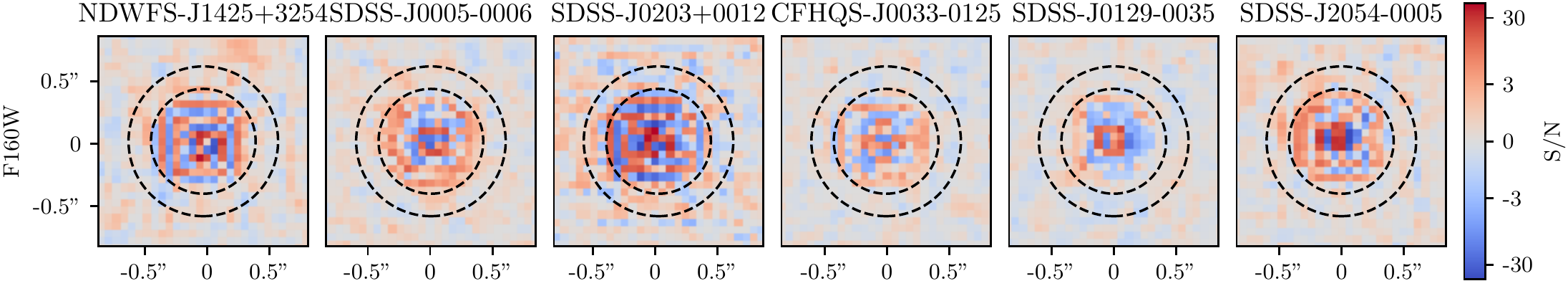}
\caption{Residual images showing the central regions of the quasars after PSF subtraction. The deep red and blue regions show pixels with formal $S/N$ with large absolute value, artifacts of the quasar subtraction technique which caused significant pixel-to-pixel variations in the residual flux. The two circles shown in each image have radii of $0\farcs42$ and $0\farcs60$, which are used when performing photometry of the underlying quasar host emission.}
\label{fig:quasar_photometry}
\end{figure*}

The formal signal-to-noise ratio ($S/N$) of the residual flux in the core of each quasar after PSF-subtraction is presented in Figure \ref{fig:quasar_photometry}.
This central region includes residual flux from the core of the quasar PSF, caused by an imperfect match of profiles of the quasar and empirical PSF used for subtraction, alongside any potential host galaxy flux. While the total flux is subtracted correctly, with a median residual consistent with zero, the pixel-to-pixel variation of the PSFs results in pixels with residual flux that is significantly larger than expected by the noise map, with formal signal-to-noise ratios of up to $\left|S/N\right|\simeq30$. In other words, the subtraction technique produces considerable residuals in the inner region.
We note that significant quasar over- or under-subtraction is unlikely as this would produce negative or positive residuals in the diffraction spikes, respectively, which are not visible.


To estimate the flux of the host without including this contaminated inner region, we instead measure the surface brightness in annuli from 7 to 10 pixels, or 0$\farcs$42--0$\farcs$60. We choose an inner radius of 7 pixels, as this is where the pixel-to-pixel variations of the $S/N$ first reach the expected/background level. This approach ignores the central core, while including enough pixels to make a reasonable detection if any flux was present.
For all quasars in both filters, no significant flux detection could be made in these annuli. The 2$\sigma$ surface brightness limits in these annuli, from the noise of each image, are given in Table \ref{tab:quasar_photometry}. 

To obtain the total magnitude limit of a given host, 
we consider a range of \mbox{S\'ersic} profiles, with a distribution of $n$ and $R_e$ guided by $z\simeq6$ observations \citep{Shibuya2015}, and take a Monte Carlo approach to determine the most likely magnitude limit given the surface brightness limit in the annulus (see Appendix \ref{Appendix}).
The 2$\sigma$ magnitude limits obtained by this method are given in Table \ref{tab:quasar_photometry}, with the J- and H-band limits ranging from 22.7--23.1 mag and 22.4--22.9 mag, respectively.

\begin{table*}
\begin{center}
\caption{Quasar Host Galaxy Detection Limits}
\begin{tabular}{c|cc|cc}
\toprule
Quasar Name & $SB_J$,~$2\sigma$ limit ($0\farcs42$-$0\farcs60$) & $m_J$, 2$\sigma$ limit (S\'ersic fit)& $SB_H$, $2\sigma$ limit ($0\farcs42$-$0\farcs60$) & $m_H$, 2$\sigma$ limit (S\'ersic fit)\\
{} &  (AB mag/$''^2$) &  (AB mag) &  (AB mag/$''^2$) &  (AB mag) \\
\hline
CFHQS-J0033-0125 &                       $24.6$ &                      $22.9$ &                       $24.4$ &                      $22.7$ \\
SDSS-J0129-0035  &                       $24.5$ &                      $22.8$ &                       $24.3$ &                      $22.6$ \\
SDSS-J0203+0012  &                       $24.4$ &                      $22.7$ &                       $24.2$ &                      $22.4$ \\
NDWFS-J1425+3254 &                       $24.7$ &                      $22.9$ &                       $24.4$ &                      $22.7$ \\
SDSS-J2054-0005  &                       $24.6$ &                      $22.9$ &                       $24.4$ &                      $22.7$ \\
SDSS-J0005-0006  &                       $24.8$ &                      $23.1$ &                       $24.6$ &                      $22.9$ \\
\hline

\end{tabular}
\label{tab:quasar_photometry}
\end{center}
\tablecomments{
Photometry of the residual image (after PSF subtraction) in the regions surrounding each quasar.
The left columns for each filter give the surface brightness in the annulus $0\farcs42$--$0\farcs60$  surrounding each quasar, as shown in Figure \ref{fig:quasar_photometry}. As no significant signal is detected, these are $2\sigma$ upper limits calculated from the noise in each image.
The right column for each filter gives magnitude limits
estimated from these surface brightness limits. These are calculated by considering a range of \mbox{S\'ersic} profiles with reasonable $n$ and $R_e$ \citep{Shibuya2015} that are constrained to have the measured surface brightness in $0\farcs42$--$0\farcs60$, and determining the most likely magnitude limit using a Monte Carlo approach (see Appendix \ref{Appendix}).
}
\end{table*}

\subsubsection{Stellar Mass Limits}
\label{StellarMassLimits}

Measuring the redshift evolution of the black hole--stellar mass relation is of key importance for understanding the co-evolution of black holes and their host galaxies. Relative to the well-studied and accurately measured local relation \citep[see, e.g., the review of][]{kormendy_2013}, at higher redshifts observations suggest that black holes are more massive compared to their hosts. 
For example, at $z\simeq1.5$ \citet{Ding2020} find a black hole--stellar mass ratio that is 2.7 times larger than the local relation, while at $z\simeq2$, \citet{peng_2006} find a black hole--bulge mass relation 3--6 times larger. At higher redshifts, existing observations of luminous $z\simeq6$ quasars with ALMA generally find black hole to dynamical mass ratios that are significantly larger than the local relation \citep[e.g.][]{Maiolino2007,Riechers2008,venemans_2012,Wang2013}.
However, many studies claim that high observed relations are a result of selection effects, \citep{Lauer2007,Schulze2011,Schulze2014,DeGraf2015,Willott2017,Ding2020}. ALMA observations of lower-luminosity $z\simeq6$ quasars indeed find these to lie on or below the local relation \citep{Willott2017,Izumi2018,Izumi2019}.

To investigate the black hole--stellar mass relation using our HST observations, we convert our host magnitude limits to limits on stellar mass. 
We calculate UV slopes $\beta$ of the hosts by fitting the relation $m=-2.5\log(\lambda^{\beta+2})+m_0$, equivalent to $f_\lambda\propto\lambda^\beta$, to the two host magnitude limits $m_J$ and $m_H$ at $\lambda=1.25$ and 1.6 $\mu$m respectively. Using this relation and the determined $\beta$ and $m_0$, we calculate the UV apparent magnitude limit $m_{\mathrm{UV}}$ as that at rest-frame 1500\AA, or $\lambda=(1+z)\times0.15\mu$m. We convert this to an absolute magnitude using $M_{\mathrm{UV}}=m_{\mathrm{UV}}-DM+2.5\log(1+z)$ where $DM$ is the distance modulus.

We adopt the $z=6$ $M_\ast-M_{\mathrm{UV}}$ relation derived by \citet{Song2016} using a large sample of galaxies from the Cosmic Assembly Near-infrared Deep Extragalactic Legacy Survey (CANDELS)/Great Observatories Origins Deep Survey (GOODS) fields and the Hubble Ultra Deep Field (HUDF):
\begin{equation}
\label{SongRelation}
\log(M_\ast) = 9.53 \pm 0.02 - (0.50 \pm 0.03)\times(M_{\mathrm{UV}}+21)
\end{equation}
with a scatter of 0.36 dex. We use a Monte Carlo technique, sampling from a uniform distribution of magnitudes ranging from $M_{\mathrm{UV}}=-20$ mag to our $2\sigma$ upper limit for each quasar host. The lower luminosity limit of $M_{\mathrm{UV}}=-20$ mag was chosen as this is as faint as high-redshift quasar hosts are expected to be from the BlueTides simulation \citep[][see Figure \ref{fig:BT_comparison}]{Feng2015,Marshall2019}. We assign a stellar mass to each sampled magnitude using Equation \ref{SongRelation}, given a normal distribution with $\sigma=0.36$ dex, to determine the resulting probability distribution of stellar masses. These stellar masses are normally-distributed, so we adopt the 2$\sigma$ upper limit from this relation as our host mass limit. We note that this results in a lower, less pessimistic limit than simply taking the $2\sigma$ upper mass limit at the $2\sigma$ magnitude limit, which is very conservative.

\begin{figure}
\centering
\includegraphics[scale=0.8]{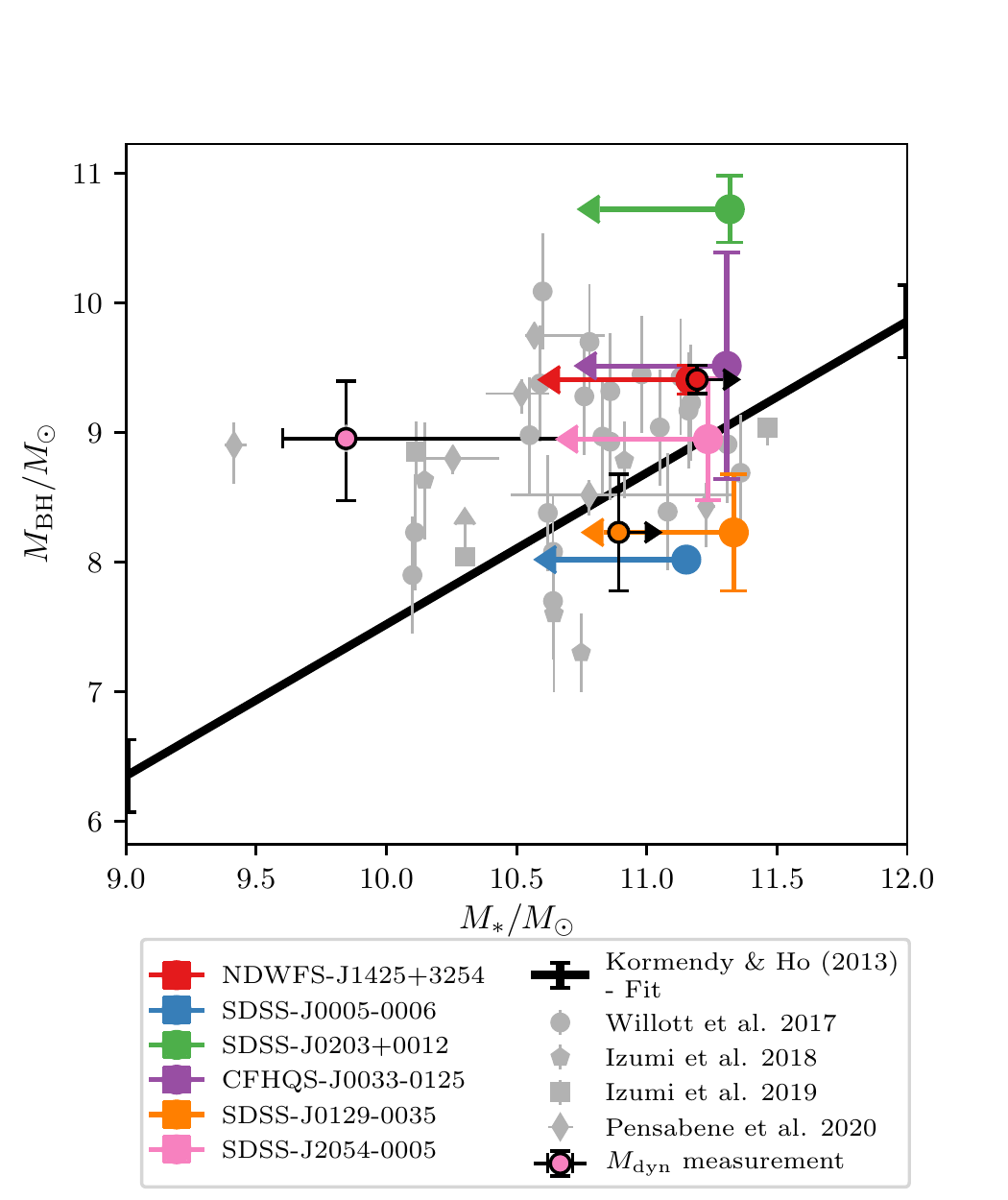}
\caption{The black hole--stellar mass relation for our $z\simeq6$ quasar host galaxies, alongside other $z\simeq6$ quasars from the literature \citep{Willott2017,Izumi2018,Izumi2019,Pensabene2020}. Existing dynamical mass measurements/limits are shown with black errorbars, for NDWFS J1425+3254 \citep[red;][]{Wang2010}, SDSS J2054-0005 \citep[pink;][]{Pensabene2020} and SDSS J0129-0035 \citep[orange;][]{Pensabene2020}.
Also shown is the $z=0$ relation of \citet{kormendy_2013}, for comparison.}
\label{fig:BHmassRelation}
\end{figure}

We present the black hole--stellar mass relation from the stellar mass limits of our $z\simeq6$ quasar hosts in Figure \ref{fig:BHmassRelation}. Our limits are consistent with the black hole--stellar mass relation from existing sub-mm observations of $z\simeq6$ quasars, \citep{Willott2017,Izumi2018,Izumi2019,Pensabene2020}, which measure the dynamical mass of the host using gas dynamics as probed by the [CII] line, and assume $M_\ast=M_{\textrm{dyn}}$.
Three of our quasars have dynamical mass measurements \citep{Wang2010,Pensabene2020}, which we also show in Figure \ref{fig:BHmassRelation}. 
Our stellar mass upper limit for SDSS J2054-0005 lies above the measured dynamical mass of $M_{\textrm{dyn}}=0.7\substack{+4.5 \\ -0.3}\times10^{10}M_\odot$ \citep{Pensabene2020}, suggesting that either our limit is significantly larger than the true stellar mass, with much deeper observations required to detect the underlying stellar emission, or the measured dynamical mass underestimates the total mass of the galaxy. 
Our stellar mass upper limit for SDSS J0129-0035 is consistent with the lower dynamical mass limit of $M_{\textrm{dyn}}>7.8\times10^{10}M_\odot$ \citep{Pensabene2020}.
The lower limit on the dynamical mass for NDWFS J1425+3254 of $M_{\textrm{dyn}}>1.56\times10^{11}M_\odot$ \citep{Wang2010} is larger than our stellar mass upper limit, suggesting that we are close to detecting the stellar component of this quasar host galaxy.

The stellar mass limits of five of our six quasars are consistent with the local \citet{kormendy_2013} $M_\ast-M_{\mathrm{BH}}$ relation. SDSS-J0203+0012, however, has a stellar mass of $M_\ast<1.89\times10^{11}M_\odot$, lower than expected by the local relation, given its extremely large black hole mass of $M_{\mathrm{BH}}=5.2\times10^{10}M_\odot$ \citep[Table \ref{tab_targets},][]{Shen2019}. However, we note that this black hole mass is determined by the C IV line, as the more robust Mg II line is not covered by ground-based observations. SDSS-J0203+0012 is a broad absorption line (BAL) quasar \citep{mortlock_2009} and so the dynamics probed by the C IV line are likely affected by the outflows. Hence we place no significance on our black hole--stellar mass relation limit for this object. 

The $M_\ast-M_{\mathrm{UV}}$ relation is derived from observations of UV-selected galaxies, and might not apply if our hosts were dusty star-forming or quiescent galaxies; as these galaxies may be particularly dusty, our mass limits would be underestimates. Mid-infrared observations to allow for detailed SED fitting, using the upcoming James Webb Space Telescope (JWST), for example, are necessary to accurately determine the stellar masses of these potentially dusty host galaxies.

\subsection{The Prevalence of Close, Blue Neighbors}
\label{the-prevalence-of-close-blue-neighbors}

\begin{table*}
\caption{Properties of Galaxies Surrounding the Quasars}
\hskip-1.82cm
\scriptsize
\begin{tabular}{llrrrrrrrrrr}
\toprule
 Quasar Name & Galaxy &             $m_J$ &           $m_J-m_H$ &   RA & Dec &  Projected Distance &         \mbox{S\'ersic} index &        $R_e$ &          $b/a$ & $M_{1500}$ &        UV slope  \\
{} &             &             (AB mag) &           (AB mag) &  & &    (kpc) & $n$      &        (kpc) &           & (AB mag) &   $\beta$     \\

\hline
CFHQS-J0033-0125 & 1$\dagger$ &  $24.7\pm0.1$ &   $0.1\pm0.1$ &   0:33:11.519 &  -1:25:23.56 &  $17.6\pm0.1$ &  $2.2\pm0.8$ &  $2.9\pm0.5$ &  $0.31\pm0.08$ &    $-22.1\pm0.2$ &  $-1.7\pm0.6$ \\
SDSS-J0129-0035 & 1* & $26.6\pm0.3$ &   $0.8\pm0.3$ &   1:29:58.088 &  -0:35:45.30 &  $18.6\pm0.2$ &  $5.5\pm1.7$ &  $1.1\pm0.4$ &  $0.64\pm0.35$ &    $-19.4\pm0.5$ &   $1.5\pm1.6$ \\
& 2* & $25.9\pm0.2$ &  $-0.7\pm0.2$ &   1:29:58.179 &  -0:35:42.23 &   $5.3\pm0.2$ &  $5.2\pm1.8$ &  $1.7\pm0.4$ &  $0.65\pm0.23$ &    $-21.4\pm0.4$ &  $-4.9\pm1.4$ \\
 & 3* &  $24.1\pm0.0$ &   $0.1\pm0.0$ &   1:29:58.114 &  -0:35:43.81 &  $10.7\pm0.1$ &  $0.5\pm0.0$ &  $1.9\pm0.1$ &  $0.46\pm0.03$ &    $-22.6\pm0.1$ &  $-1.6\pm0.2$ \\
SDSS-J0203+0012 
& 1a* &  $23.3\pm0.1$ &   $0.1\pm0.1$ &   2:03:31.865 &   0:12:25.09 &  $21.3\pm0.1$ &  $0.5\pm0.0$ &  $1.8\pm0.0$ &  $0.22\pm0.02$ &   $-23.8\pm0.3$ &  $-1.7\pm1.3$   \\
& 1b* &  $24.1\pm0.1$ &  $-0.1\pm0.3$ &   2:03:31.860 &   0:12:24.98 &  $22.1\pm0.1$ &  $2.1\pm0.7$ &  $1.6\pm0.3$ &  $0.76\pm0.20$ &   - &    -   \\
& 2 &  $26.6\pm0.2$ &   $0.2\pm0.3$ &   2:03:31.883 &   0:12:28.62 &  $15.9\pm0.3$ &  $4.2\pm2.0$ &  $1.6\pm0.4$ &  $0.57\pm0.29$ &    $-19.9\pm0.4$ &  $-1.1\pm1.6$  \\
& 3 &  $26.2\pm0.3$ &   $0.2\pm0.3$ &   2:03:32.180 &   0:12:25.94 &  $16.0\pm0.2$ &  $4.7\pm1.9$ &  $1.8\pm0.5$ &  $0.59\pm0.31$ &    $-20.3\pm0.5$ &  $-1.2\pm1.6$  \\
NDWFS-J1425+3254 & 1$\dagger$ &  $24.6\pm0.1$ &   $0.4\pm0.1$ &  14:25:16.767 &  32:54:06.98 &   $8.4\pm0.1$ &  $3.6\pm0.7$ &  $2.6\pm0.4$ &  $0.81\pm0.21$ &    $-21.8\pm0.2$ &  $-0.4\pm0.8$  \\
 &  2* &  $24.4\pm0.1$ &   $0.1\pm0.1$ &  14:25:16.733 &  32:54:05.49 &   $3.4\pm0.2$ &  $0.5\pm0.0$ &  $2.7\pm0.1$ &  $0.94\pm0.07$ &    $-22.3\pm0.2$ &  $-1.6\pm0.6$  \\
SDSS-J2054-0005 & 1* & $23.7\pm0.1$ &  $0.0\pm0.1$&  20:54:06.075 &  -0:05:18.12  &  $12.3\pm0.1$ &  $1.7\pm0.2$ &  $4.2\pm0.2$ &  $0.39\pm0.04$ &    $-23.1\pm0.1$ &  $-2.2\pm0.4$ \\
& 2* &  $24.3\pm0.1$ &   $0.4\pm0.1$ &  20:54:06.028 &  -0:05:17.48 &  $15.6\pm0.1$ &  $0.9\pm0.2$ &  $2.2\pm0.2$ &  $0.49\pm0.05$ &    $-22.1\pm0.1$ &  $-0.1\pm0.4$ \\
& 3* &  $23.3\pm0.0$ &   $0.1\pm0.0$ &  20:54:06.080 &  -0:05:17.31 &  $11.1\pm0.0$ &  $1.6\pm0.1$ &  $0.7\pm0.0$ &  $0.81\pm0.03$ &    $-23.5\pm0.0$ &  $-1.6\pm0.1$ \\
& 4* &  $24.8\pm0.2$ &  $-0.6\pm0.1$ &  20:54:05.998 &  -0:05:15.14 &  $22.7\pm0.0$ &  $5.7\pm1.0$ &  $2.3\pm0.4$ &  $0.41\pm0.15$ &    $-22.5\pm0.3$ &  $-4.6\pm0.9$  \\
& 5$\dagger$ &  $24.9\pm0.1$ &   $0.2\pm0.1$ &  20:54:06.265 &  -0:05:19.87 &  $15.7\pm0.0$ &  $7.0\pm0.8$ &  $2.2\pm0.5$ &  $0.63\pm0.24$ &    $-21.7\pm0.2$ &  $-1.1\pm0.8$ \\
& 6$\dagger$ &  $25.1\pm0.1$ &   $0.4\pm0.1$ &  20:54:06.376 &  -0:05:19.41 &  $19.4\pm0.1$ &  $4.6\pm0.8$ &  $3.7\pm0.4$ &  $0.72\pm0.13$ &    $-21.4\pm0.2$ &  $-0.2\pm0.8$ \\
& 7$\dagger$ &  $25.3\pm0.2$ &   $0.4\pm0.2$ &  20:54:06.336 &  -0:05:18.94 &  $14.8\pm0.0$ &  $5.6\pm0.9$ &  $1.7\pm0.2$ &  $0.55\pm0.11$ &    $-21.1\pm0.3$ &  $-0.2\pm1.1$ \\
& 8 &  $26.0\pm0.2$ &   $0.4\pm0.2$ &  20:54:06.334 &  -0:05:19.44 &  $16.7\pm0.0$ &  $4.8\pm0.8$ &  $2.2\pm0.4$ &  $0.56\pm0.16$ &    $-20.5\pm0.3$ &  $-0.5\pm1.1$ \\
SDSS-J0005-0006 
 &  1a*&  $25.1\pm0.1$ &  $-0.3\pm0.3$ &   0:05:52.046 &  -0:06:59.94 &  $10.8\pm0.1$ &  $0.7\pm0.2$ &  $1.0\pm0.1$ &  $0.85\pm0.10$ &    $-22.1\pm0.6$ &  $-1.8\pm2.1$ \\
 & 1b*& $25.7\pm0.3$ &   $0.5\pm0.2$ &   0:05:52.038 &  -0:06:59.83 &   $9.8\pm0.2$ &  $6.4\pm1.2$ &  $2.0\pm0.9$ &  $0.59\pm0.45$ &      - &    - \\
& 2* &  $25.7\pm0.1$ &   $0.7\pm0.4$ &   0:05:52.217 &  -0:06:56.57 &  $23.5\pm0.1$ &  $5.7\pm1.5$ &  $2.7\pm0.6$ &  $0.41\pm0.16$ &    $-20.4\pm0.4$ &   $1.0\pm1.7$  \\
& 3$\dagger$ &  $25.4\pm0.1$ &   $0.0\pm0.2$ &   0:05:52.032 &  -0:06:55.61 &  $17.3\pm0.1$ &  $2.1\pm0.9$ &  $2.6\pm0.5$ &  $0.33\pm0.11$ &    $-21.3\pm0.3$ &  $-2.0\pm0.9$ \\
\hline
\end{tabular}
\label{tab:galaxy_properties}
\tablecomments{Properties of the galaxies surrounding each of the quasars, with the central value denoting the maximum posterior model and the error extracted from the MCMC fits. Galaxies with an a/b marked next to their identifier (column 2) are those which needed two \mbox{S\'ersic} profiles to be fit to match their brightness distribution. The properties for the individual fits are given, but their UV magnitude and slope $\beta$ are calculated by combining both \mbox{S\'ersic} magnitudes. Asterisks (*) denote companions with UV magnitudes and/or slopes that make them unlikely to be $z\simeq6$ galaxies, and are instead likely to be foreground interlopers. Daggers ($\dagger$) denote potential $z\simeq6$ companions with UV magnitudes brighter than $M^\ast$ at $z=6$.}
\end{table*}

\begin{figure*}
\centering
\includegraphics[scale=0.8]{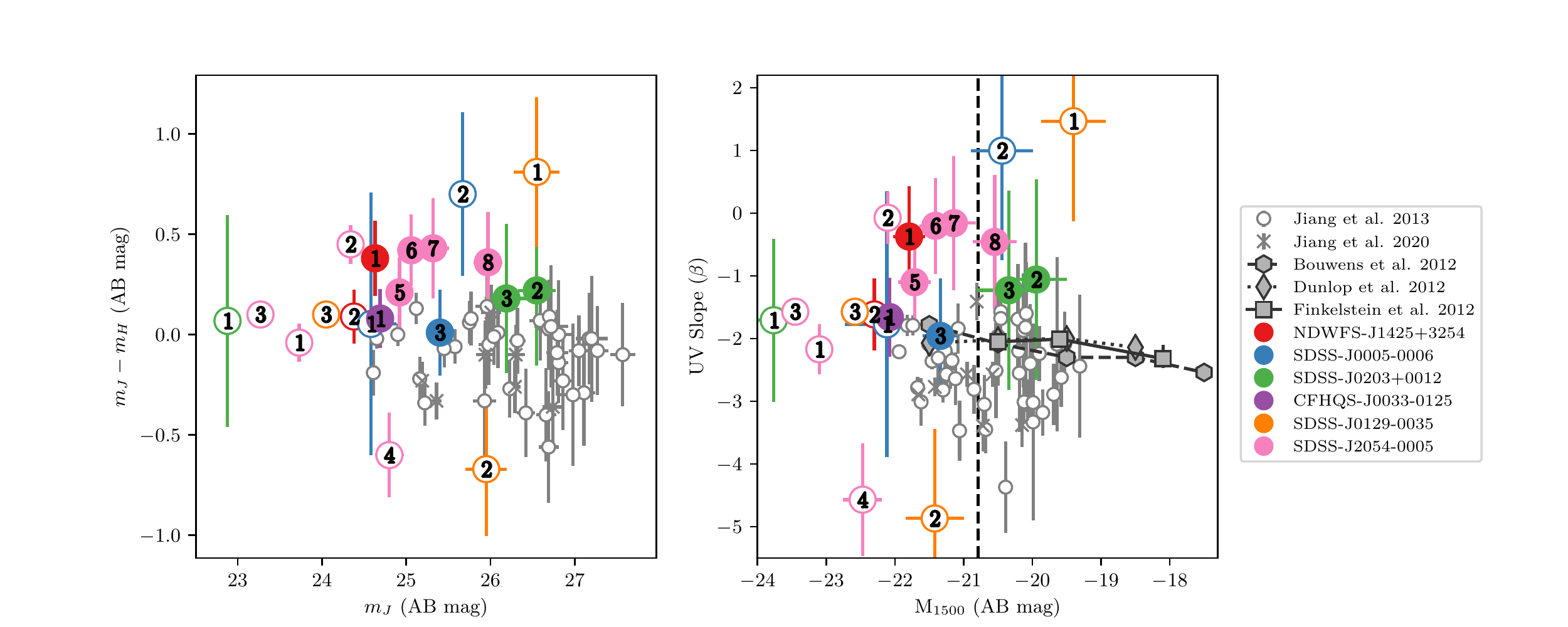}
\caption{
Left: J-H color vs. J-band magnitude and Right: Rest-frame UV slope $\beta$ vs. 1500~\AA{} absolute magnitude, measured for each of the galaxies within $\simeq3''$ of our six quasars, assuming they are at the same redshift as the quasar.
Filled colored circles show those with colors and magnitudes consistent with $z\sim6$ galaxies, while open colored circles show candidates which are likely to be foreground interlopers given their colors and magnitudes. The numerical labels correspond to labels in the individual quasar images (Figures \ref{fig:quasar1_resids}-\ref{fig:quasar3_resids}), and Table \ref{tab:galaxy_properties}, for ease of comparison.
Grey symbols represent  spectroscopically-confirmed $z\simeq6$ galaxies from \citet{Jiang2013,Jiang2020} and the average relations for dropout-selected LBGs from \citet{Bouwens2012}, \citet{Dunlop2012}, and \citet{Finkelstein2012} (see legend). 
The dashed black line shows the value of $M^*$ \citep{Finkelstein2016}.}
\label{fig:colour_mag}
\end{figure*}

Figures \ref{fig:quasar1_resids}--\ref{fig:quasar3_resids} reveal that all of our six quasars have neighboring galaxies within the surrounding $6\farcs5\times6\farcs5$. For
SDSS-J0129-0035, NDWFS-J1425+3254 and SDSS-J0005-0006, some companions overlap with the quasar PSF, highlighting the need for the quasar PSF subtraction in order to fully understand the local quasar environment.

The properties of these 20 neighboring galaxies from the maximum posterior model found by \soft{psfMC} are listed in Table \ref{tab:galaxy_properties}. We calculate their UV magnitudes and slopes following the same procedure as for the host galaxies (see Section \ref{StellarMassLimits}). The magnitudes and colors of these galaxies are displayed in Figure \ref{fig:colour_mag}, along with samples of star-forming galaxies at $z\simeq6$. Four of these companions have colors/UV-slopes that are too red ($\beta>0$) or too blue ($\beta<-4$) to be consistent with $z\simeq6$ galaxies. In addition, seven galaxies are too bright
to be likely at this redshift, with magnitudes brighter than $M_{\mathrm{UV}}=-22.1$ mag, the magnitude of the brightest spectroscopically-confirmed $z\simeq6$ galaxy in the sample of \citet{Finkelstein2015}. The remaining 9 companion galaxies---surrounding five of our six quasars---have UV magnitudes and slopes consistent with those of star-forming galaxies at $z\simeq6$. The majority of these are brighter than $M^\ast$ \citep[$-20.79$ mag at $z=6$,][see Table \ref{tab:galaxy_properties}]{Finkelstein2016}. Unfortunately, existing observations at sub-mm to radio wavelengths do not resolve and/or detect the individual sources \citep[e.g.][]{Wang2013}, so morphological comparisons with existing data are not possible. 

These 9 potential companion galaxies are separated from the quasars by $1\farcs4$--$3\farcs2$, corresponding to projected distances of 8.4--19.4 kpc.
Simulations show that galaxies with companions at similar separations have higher AGN fractions \citep{McAlpine2020}, and also enhanced star formation rates \citep{Patton2020}. This suggests that these companions, if their true 3D distance is of order their projected distance, could be interacting with the quasar host galaxies and may potentially have triggered or enhanced the observed AGN activity.
However, tidal features from any such interaction would likely be rendered invisible due to the $(1+z)^4$ surface brightness dimming at $z\simeq6$.

\begin{figure*}
\centering
\includegraphics[scale=0.8]{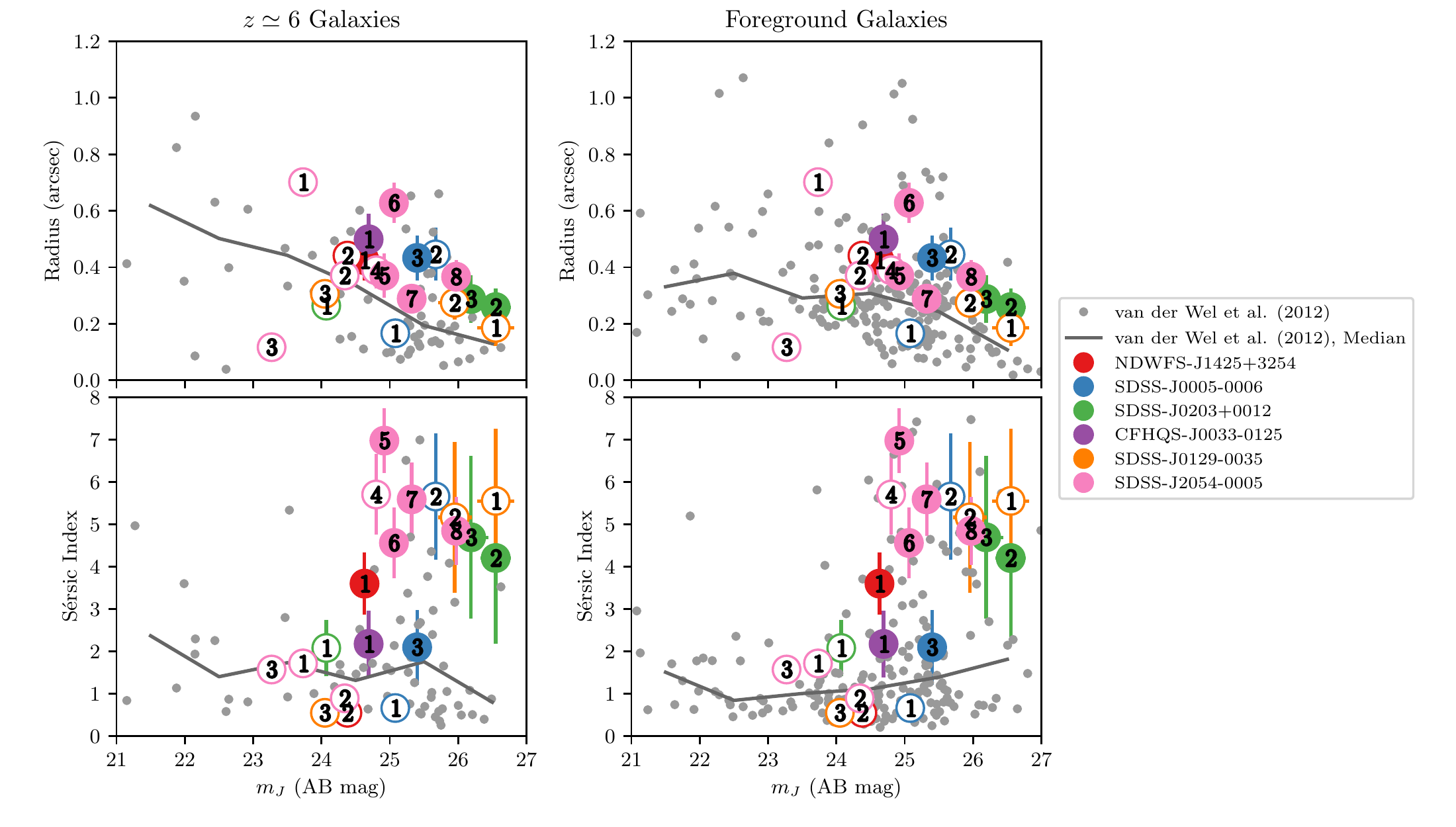}
\caption{The morphological properties measured for each of the galaxies within $3''$ of our 6 quasars.
Top row: effective radius $R_e$ vs. J-band magnitude.
Bottom row: S\'ersic index $n$ vs. J-band magnitude.
Filled colored circles show those with colors and magnitudes consistent with $z\simeq6$ galaxies, while open colored circles show candidates which are likely to be foreground interlopers given their colors and magnitudes. The numerical labels correspond to labels in the individual quasar images (Figures \ref{fig:quasar1_resids}-\ref{fig:quasar3_resids}), and Table \ref{tab:galaxy_properties}, for ease of comparison.
Also shown are measurements of galaxy sizes in the CANDELS GOODS-South survey from \citet{VanDerWel2012}, for comparison. 
Left panel: \citet{VanDerWel2012} galaxies with best estimate redshift $z>5.5$, and a 95 per cent confidence that $z>5$.
Right panel: Galaxies within $3''$ of these $z>5.5$ galaxies, at any redshift. Grey dots show individual galaxies, and the grey line shows the median in bins of 1 magnitude.
}
\label{fig:3DHST}
\end{figure*}

We examine the relationship between size, S\'ersic index and magnitude for these neighboring galaxies in Figure \ref{fig:3DHST}, in comparison to $z\simeq6$ galaxies in the CANDELS GOODS-South sample \citep{VanDerWel2012,Brammer2012,Skelton_2014,Momcheva2016}.
This shows that our companion galaxy sizes are as large as the largest $z\simeq6$ CANDELS field galaxies, but are larger than the median size of $z\simeq6$ field galaxies by an average of $0\farcs1$.
Their S\'ersic profiles are as steep as the steepest S\'ersic indexes of CANDELS $z\simeq6$ field galaxies, but are larger than the median S\'ersic index of $z\simeq6$ field galaxies by an average of 2. Thus, our potential quasar companion galaxies have morphological parameters that are consistent with the larger and steeper-S\'ersic CANDELS $z\simeq6$ field galaxies. 
Similar conclusions are made when comparing our potential companion galaxies to neighbors within $3''$ of $z\simeq6$ galaxies in the CANDELS GOODS-South sample \citep[Figure \ref{fig:3DHST},][]{VanDerWel2012}, of which the majority (95\%) are foreground objects at $z<5.5$. Our potential companions have sizes and S\'ersic indices that are larger than the median of the neighbors of $z\simeq6$ CANDELS galaxies, although their properties are reasonably consistent with the more massive neighbors.
Hence, based on their size and S\'ersic distributions we cannot determine whether our observed neighbors are more likely to be $z\simeq6$ galaxies than foreground interlopers.

\begin{figure}
\centering
\includegraphics[scale=0.8]{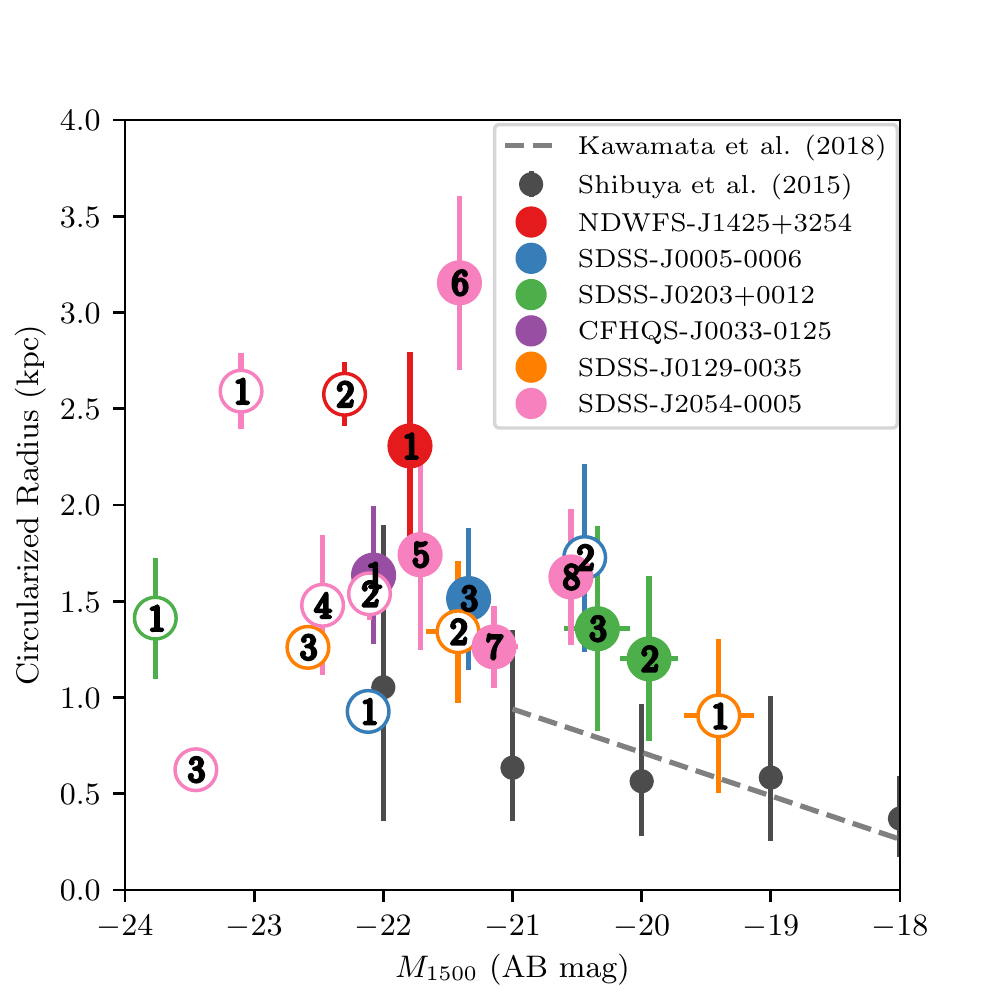}
\caption{The circularized effective radius $R_e \sqrt{b/a}$ vs. 1500~\AA{} absolute magnitude measured for each of the galaxies within $3''$ of our 6 quasars, assuming they are at the same redshift as the quasar.
Filled colored circles show those with colors and magnitudes consistent with $z\sim6$ galaxies, while open colored circles show candidates which are likely to be foreground interlopers given their colors and magnitudes. The numerical labels correspond to labels in the individual quasar images (Figures \ref{fig:quasar1_resids}-\ref{fig:quasar3_resids}), and Table \ref{tab:galaxy_properties}, for ease of comparison. The size--luminosity relations from the high-redshift observations of \citet{Shibuya2015} and \citet{Kawamata2018} are also shown, for comparison. }
\label{fig:size_mag}
\end{figure}

We show the relationship between size and UV absolute magnitude for these potential companion galaxies in Figure \ref{fig:size_mag}, assuming they are at the same redshift as the quasar. Our objects that have UV magnitudes and slopes consistent with $z\simeq6$ galaxies lie on a relatively tight size-luminosity relation. 
This relation is fairly consistent with, but somewhat higher than, that of $z\simeq6$ Lyman-break galaxies measured by \citet{Shibuya2015} (for galaxies with $-22\lesssim M_{\textrm{UV}} \lesssim -18$ mag) and \citet{Kawamata2018} (for galaxies with $-21.5\lesssim M_{\textrm{UV}} \lesssim -12$ mag), with our objects having larger sizes at the same luminosities. 

We note that the measured sizes of galaxies may be affected by systematic differences between these studies. 
For example, the treatment of the sky background in the fitting can have a significant impact on the resulting S\'ersic fit parameters \citep[see, e.g.,][]{Guo2009,Bruce2012}. We include the background level as a free parameter in the MCMC-fitting, which can result in larger values of $R_e$ and S\'ersic index $n$ than if the background level is fixed \citep{Bruce2012}, potentially contributing to some of the discrepancy between our results and those of \citet{Shibuya2015}, who assume a fixed background level.
\citet{Bruce2012} show that there is a positive correlation between measured $R_e$ and S\'ersic index $n$. \citet{Shibuya2015} assume $n=1.5$ for Lyman-break galaxies, and \citet{Kawamata2018} fix $n=1$. As our fitting method finds larger S\'ersic indices for our potential $z\simeq6$ companion galaxies, with a mean of $n=4.3$, this correlation may explain, at least in part, our larger measured sizes.
\\

Using the number of galaxies observed in HST data of comparable depth \citep[WFC3 ERS2 field, $m_H<26.5$;][see their figure 12]{Windhorst2011}, the average number of galaxies observed is $\approx373,000$ per square degree, or 0.0288 objects per square arcseconds, {\it if} galaxies are uniformly distributed on the sky. We thus expect to find on average 7.3 {\it random} foreground objects within our six $\approx 6\farcs5 \times 6\farcs5$ quasar images (total area $\approx250$ square arcseconds; Figures \ref{fig:quasar1_resids}--\ref{fig:quasar3_resids}), at $z\ll6$ and unrelated to our quasars. 
The surface density variations in these numbers due to foreground cosmic variance and photometric zeropoint errors is expected to be $\lesssim$15\% in the H-band \citep[see, e.g., figure 3 of ][]{Driver2016}, so we would on average expect $\lesssim$ 8.4 random foreground objects at $z\ll6$.

Assuming that the galaxy neighbor distribution follows a Poisson distribution, the probability of observing a total of 20 galaxies within the $\approx250$ square arcsecond area, $4\sigma$ above the expected value of 8.4 foreground galaxies, is $\lesssim0.0003$. Hence, given the unlikelihood that so many galaxies would be found by chance, it is probable that some of these objects are physically associated with the quasars. 
Of the 20 close neighbors, 11 have UV magnitudes and slopes that suggest that they are unlikely to be at $z\simeq6$. Finding these 11 foreground galaxies is consistent within $1\sigma$ of these Poisson expectations. The remaining 9 neighbors have UV magnitudes and slopes consistent with known $z\simeq6$ galaxies. We will adopt this number of 9 potential quasar companion galaxies in our discussion below, however we note that the true number of $m_J<26.5$ $z\simeq6$ companion galaxies in our images could be between 0 and 20, with 9 a more reasonable upper limit on the expected number based on the above arguments. 

\begin{figure}
\centering
\includegraphics[width=\linewidth]{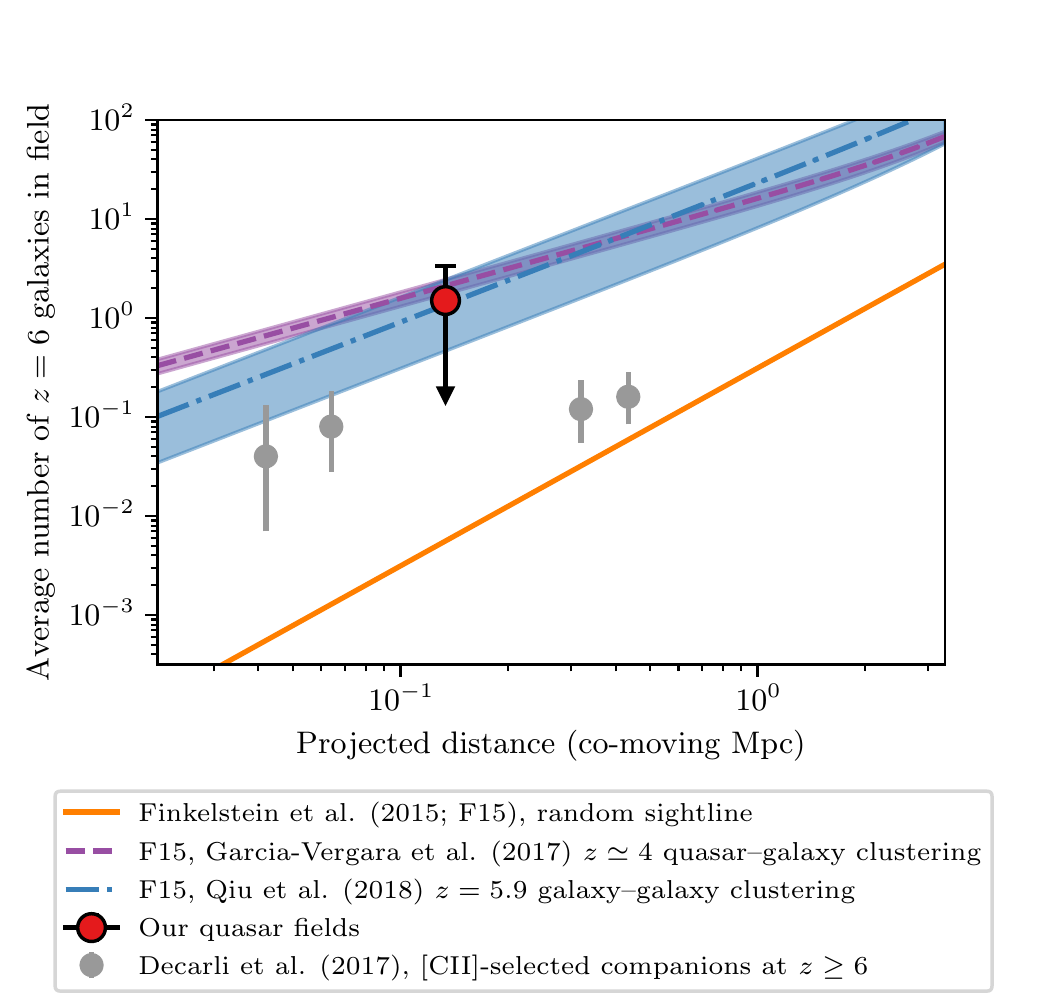}
\caption{The average number of potential $z\simeq6$ companion galaxies found within $3\farcs2$ (0.13 cMpc) of our six quasars, compared with expectations for counts of $z=6$ galaxies in random fields from the \citet{Finkelstein2015} luminosity function, as a function of projected distance from the quasar. 
We consider a cylindrical volume centred on the quasar with depth $\Delta z = 1$. Errors on our observation show the range of 1--20 true $z\simeq6$ companions, around our best estimate of 9, where 20 is the total number of objects found within $3\farcs2$ of our six quasars; the lower error is marked with an arrow to account for the (unlikely) limiting case that none of the neighboring galaxies are true $z\simeq6$ companions. 
Also shown are the observations of $z\geq6$ quasar companions of \citet{Decarli2017}, for comparison.
We also plot the \citet{Finkelstein2015} luminosity function predictions modified to account for the effect of large-scale clustering. We take two models for the excess in the galaxy number density $\xi(r)=(r_0/r)^\gamma$, with  $r_0=8.83\substack{+1.39 \\-1.51} h^{-1} $ cMpc  and $\gamma=2.0$ from the quasar--Lyman break galaxy clustering of $z=4$ galaxies measured by \citet{Garcia2017}, and $r_0=5.3\substack{+2.3\\-2.6} h^{-1} $ cMpc  and $\gamma=1.6$ from the galaxy--galaxy clustering measurements of $z=5.9$ $M_{UV}<-19.99$ galaxies \citep{Qiu2018}. 
}
\label{fig:overdensity}
\end{figure}

In Figure \ref{fig:overdensity} we plot the average  number  of potential $z\simeq6$  companion  galaxies  found  in our quasar fields, compared to expectations for number counts of $z=6$ galaxies in random fields \citep{Finkelstein2015}. We see significant excess compared to expectations of random pointings, with our average of $\sim$1.5 galaxies within 0.13 cMpc of the quasars ($3\farcs2$ at $z=6$) higher than the expected number of 0.0056 galaxies, corresponding to an overdensity of a factor of $\sim$270.
This large excess is similar to the overdensity found by \citet[][see their figure 3b]{Decarli2017}, who detected 4 companion galaxies in [CII] with ALMA around 4 of 25 $z\gtrsim6$ quasars. 

\citet{Decarli2017} find that their measured overdensity is consistent with measurements of quasar--Lyman break galaxy clustering at $z\simeq4$ \citep{Garcia2017}, when applied to the [CII] luminosity function. 
We consider the models of the $z\simeq4$ \citet{Garcia2017} quasar--galaxy clustering and the $z=5.9$ galaxy--galaxy clustering of \citet{Qiu2018} to account for this effect, and find that our measurement is consistent with both clustering models applied to the \citet{Finkelstein2015} luminosity function. Thus, as with the \citet{Decarli2017} sample, our observed potential $z\simeq6$ companion counts can be explained by expectations for high-redshift galaxy clustering. Hence, while we find that our quasars are in environments that are more dense than the average field density, with an overdensity of a factor of $\sim$270, they are in similar environments to that expected for a typical \emph{luminous} $z\simeq6$ field galaxy. In fact, if we have overestimated the number of true companion galaxies at 9, then our quasars may be in somewhat less dense regions than the typical luminous $z\simeq6$ galaxy

While \citet{Decarli2017} had a secure number of $z\simeq6$ companions from
ALMA [CII] redshifts, we present a consistent upper limit from the number of
possible $z\simeq$6 companions following the above arguments. Clearly, JWST integral field redshifts, ALMA [CII] redshifts, or VLT MUSE redshifts would be needed to determine the real number of companion galaxies around our quasars, and their overdensity compared to the field at $z\simeq$6. 

\subsubsection{Additional Observations of NDWFS-J1425+3254}
The quasar NDWFS-J1425+3254 shows further evidence for having close companions.
The discovery spectrum of  \citet{Cool2006} shows a significant absorption feature at roughly $8350$~\AA,~ $20$~\AA~ red-ward of Lyman-$\alpha$. This line could potentially be caused by HI absorption from a companion galaxy infalling at $720\textrm{~km~s}^{-1}$ \citep{Mechtley2014}. By assuming the system is virialized and that the companion is at a projected distance of 4.8 kpc, \citet{Mechtley2014} find that this corresponds to a dynamical mass of $\sim5.8\times10^{11}M_\odot$.
Using the \citet{Song2016} $M_\ast-M_{\mathrm{UV}}$ relation (Equation \ref{SongRelation}), from our detection limits the host of NDWFS-J1425+3254 has a stellar mass of $M_\ast<2.0\times10^{11}M_\odot$, with the two companions having masses of $\sim7.8\times10^8M_\odot$ and $\sim1.3\times10^9M_\odot$. The properties of the CO~$(6-5)$ line \citep{Wang2010} provide independent evidence for a group-like gravitational potential; the line fit gives a FWHM of $690 \pm 180$~km~s$^{-1}$, and the peak of the emission is redshifted (\zeq{5.89}) from the reported Ly$\alpha$ redshift \citep[\zeq{5.85},][]{Cool2006}.

\begin{figure*}
\centering
\includegraphics[width=0.9\textwidth]{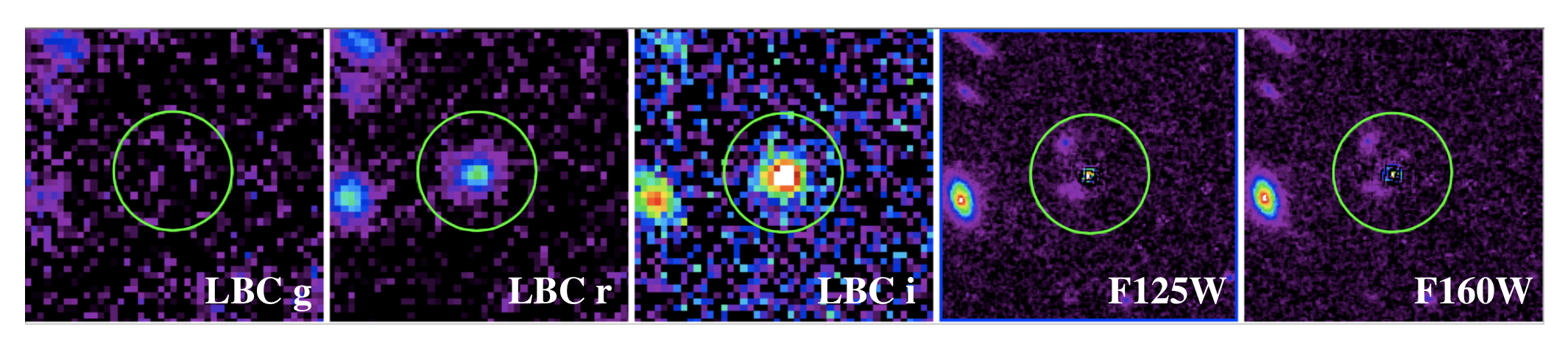}
\caption{Thumbnail images of NDWFS-J1425+3254. From left to right: LBT/LBC $g$-, $r$-, and $i$-bands, and HST WFC3 IR J- and H-bands. The companion galaxies are visible within the green circle of radius $2\farcs0$ in the point source-subtracted WFC3 IR images. Lyman-$\alpha$ emission from the quasar at $\simeq8330$\AA~ is captured by the $i$-band image, while
Lyman-Werner flux from the quasar is bright enough to be seen in the $r$-band, even in these LBT observations \citep[see the discovery spectrum of][]{Cool2006}. Due to the seeing of the ground-based images, estimated as $\simeq0\farcs8-1\farcs0$ FWHM, point source subtraction on the $r$-band LBC image produces an inconclusive upper limit for the combined companion galaxy flux. }
\label{fig:multiband}
\end{figure*}

To further investigate this system, we obtained observations with the Large Binocular Camera (LBC) on the Large Binocular Telescope (LBT) in the $g$-, $r$-, and $i$-bands (Fig \ref{fig:multiband}). No LBC $g$-band flux is detected in a $2\farcs0$ aperture to a limit of  $m_g\gtrsim28.3$~mag, with Lyman-Werner flux from the quasar detected in the $r$-band at $m_r=24.7$~mag. Even in decent seeing conditions ($\approx 0\farcs8-1\farcs0$) the ground-based PSF of the quasar has broad wings that significantly affect the detection limit of close companions out to 2\farcs0 or greater \citep[see, e.g.,][]{Ashcraft2018}. A best-effort point-source subtraction results in an upper limit of $m_r\gtrsim25.7$~mag for the more distant of the two companions. The J- and H-band detections but faint $g$- and $r$-band limits are sufficient to exclude the possibility that these companions are blue foreground galaxies, but not that they could be red luminous galaxies at $z\simeq1.1$. Additional observations are therefore required to confirm that these `companion' galaxies are indeed at $z\simeq6$, and not foreground interlopers.

\section{Discussion}\label{discussion}

\subsection{The Dust Content of Quasar Hosts}
\label{sec:Dust}

\begin{figure*}
\centering
\begin{minipage}[b]{0.49\textwidth}
\includegraphics[width=0.98\linewidth]{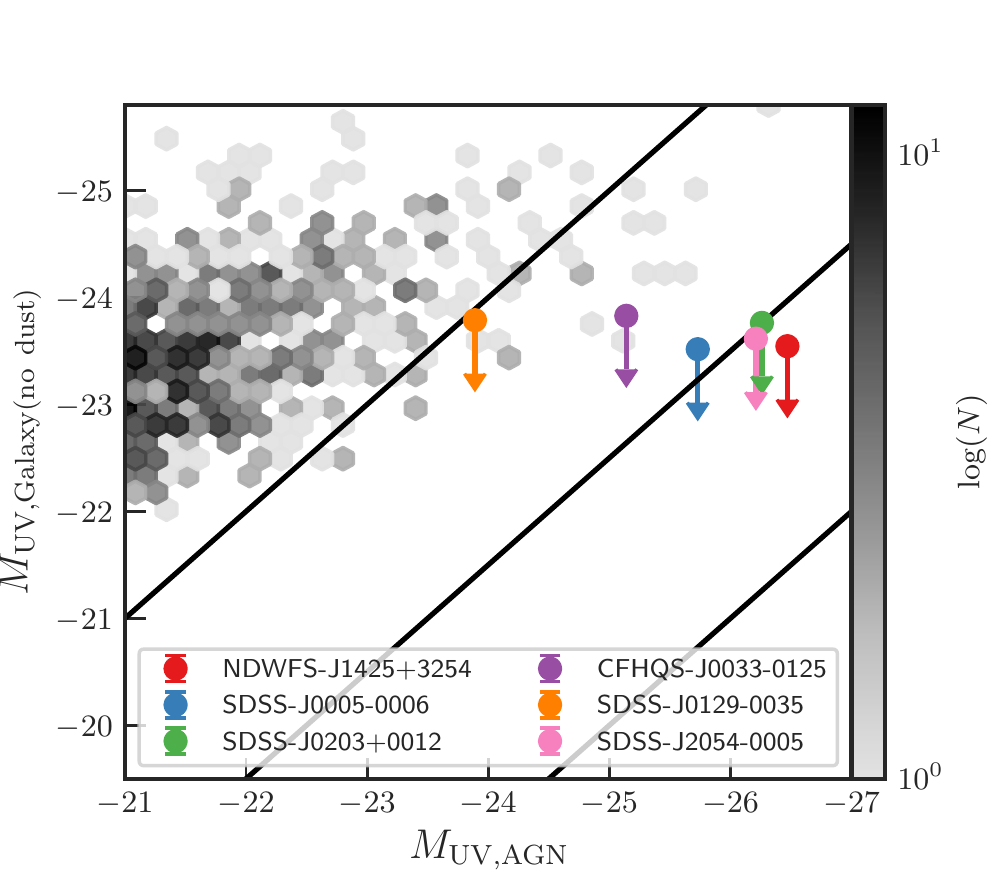}
\end{minipage}
\begin{minipage}[b]{0.49\textwidth}
\includegraphics[width=0.98\linewidth]{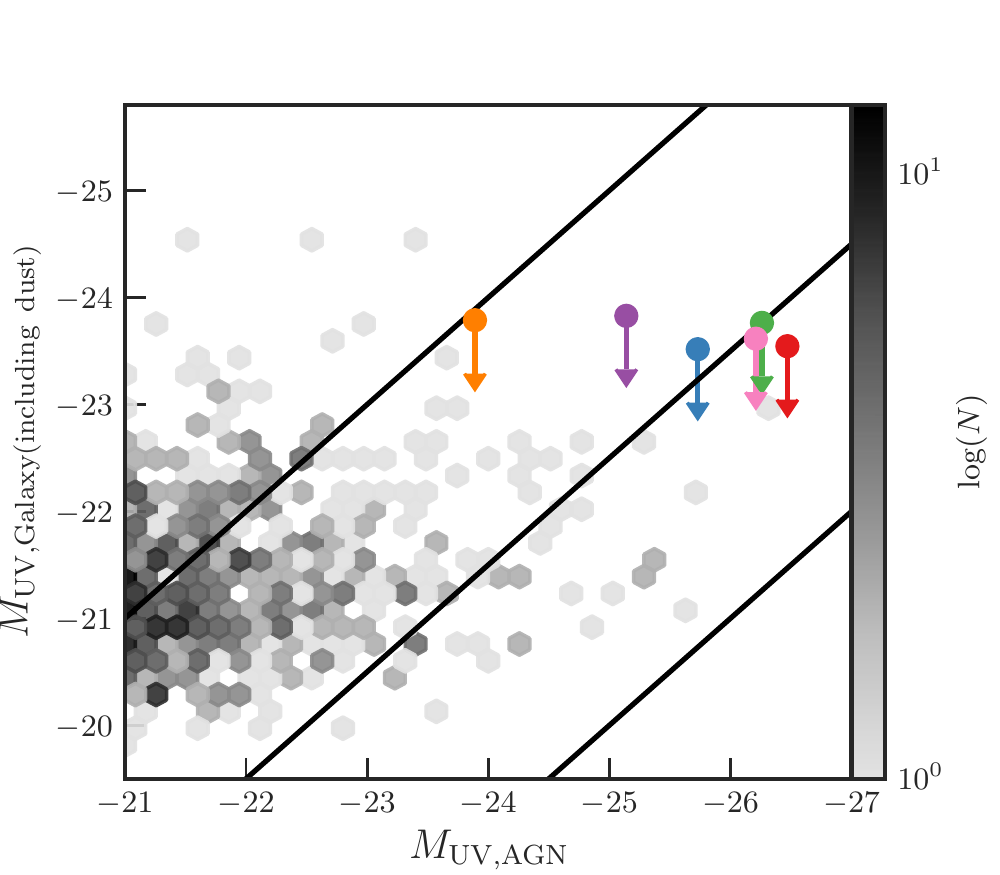}
\end{minipage}
\caption{The relation between quasar and host galaxy UV-luminosity for our sample (colored upper limits; see legend) and for the simulated $z=7$ quasars from the BlueTides simulation (grey density plots). The left panel shows the BlueTides host galaxies' intrinsic UV-luminosities, while the right panel shows them after dust attenuation, which is calculated using the density of gas in the simulation \citep[see][]{Ni2019,Marshall2019}. Diagonal black lines show where the ratio of quasar to host brightness is 1:1, 10:1 and 100:1.}
\label{fig:BT_comparison}
\end{figure*}

To understand the magnitude limits and dust properties of our quasar host galaxies, we consider the sample of $z=7$ quasars in the BlueTides simulation \citep{Feng2015}. BlueTides is a large-scale cosmological hydrodynamical simulation, which models the evolution of $2\times 7040^{3}$ particles in a cosmological box of volume $(400/h ~\rm{cMpc})^3$ from initial conditions at $z=99$ to $z=7$, the lowest published redshift to date \citep{Marshall2019,Ni2019}.
Figure \ref{fig:BT_comparison} presents the relation between galaxy and quasar UV luminosity, for our observations and the BlueTides quasars. 
Both the intrinsic and dust-attenuated galaxy magnitudes for the BlueTides galaxies are shown, with the dust attenuation of galaxies modeled in BlueTides using the density of metals along a line of sight \citep[for full details, see][]{Marshall2019}. From the BlueTides simulation, we find that hosts of $M_{\mathrm{UV}}<-23$ mag quasars at $z=7$ have between 1.4 to 3.8 mag extinction in the UV, with an average of $A_{\mathrm{UV}}=2.6$ mag, corresponding to $A_{\mathrm{V}}=1.0$ \citep{Calzetti2000}.

From Figure \ref{fig:BT_comparison}, we see that the majority of intrinsic UV magnitudes for host galaxies of similar luminosity quasars in BlueTides are brighter than our host galaxy upper limits, with only a small percentage consistent with our limits. Given that we make no detections of all six quasar hosts, our upper limits are sufficient to rule out the possibility that our quasars are generally hosted by dust-free galaxies. Instead, our limits favour host galaxies with significant dust-attenuation, consistent with the $\langle A_{\mathrm{UV}}\rangle=2.6$ mag that is seen for the BlueTides quasar hosts \citep[Figure \ref{fig:BT_comparison},][]{Marshall2019}. 
If the BlueTides sample including dust obscuration is representative of the true $z\simeq6$ quasar population, our upper limits are brighter than the host magnitudes that are expected, and future observations would need to probe at least $\sim1$ mag fainter to begin to detect the host emission. We will focus on integrating BlueTides with our observational techniques to make specific predictions for upcoming JWST observations in future work.

SDSS J0005-0006 was found to be a dust-poor quasar by \citet{Jiang2013}, as it was undetected with the Spitzer Space Telescope at 15.6 and 24 $\mu$m. 
Further observations by \citet{leipski_2014} detected the quasar with Spitzer at these wavelengths, however did not detect emission at $\geq 100 \mu$m with Herschel, and so they also conclude that the quasar is deficient of hot dust compared to the majority of quasars in their sample.
Non-detection of the quasar with the Max Planck Millimeter Bolometer Array (MAMBO) at 250 GHz \citep{Wang2008} results in upper limits of the dust mass of the host of $M_{\textrm{dust}}<1.9\times10^8 M_\odot$ \citep{calura_2014}.  While these observations do not detect significant amounts of dust in this system, it is still possible for some dust to be present, resulting in low-level dust attenuation of the host galaxy.
Thus, while our magnitude limit for the host of SDSS J0005-0006 is fainter than the magnitudes expected of quasar hosts with no dust attenuation from the BlueTides simulation, our non-detection of the host can be reasonably explained by some minor dust attenuation in the generally dust-deficient system. Additionally, if the simulation was run to $z\simeq6$, we would expect to see a larger sample of luminous quasars, and potentially more with host luminosities fainter than our magnitude limits, which could explain our observations. Thus, the `dust-free' nature of SDSS J0005-0006 is not in significant tension with our overall dust predictions.
\\

Our six quasars were selected in the $z$-band as $i$-band dropouts with rest-frame UV luminosities at $z\simeq 6$ of $-26.5\lesssim M_{UV} \lesssim -24$ mag (Table \ref{tab_targets}). Hence, their UV accretion disks are still, by selection, remarkably well visible. Five of our $z\simeq6$ quasars were also selected to have significant FIR emission, and as a consequence the young stellar populations in their host galaxies are not visible in the best high dynamic range J- and H-band images that HST can produce. 

We inferred that their host galaxies are likely considerably dusty ($\langle A_{\mathrm{UV}}\rangle=2.6$ mag), yet the embedded quasars are easily detected in the rest-frame UV and thus not significantly obscured \citep[see, e.g.,][]{Vito2019}. This must have significant consequences for the geometry of the small and large scale dust distribution. One possible explanation is that the embedded rapidly-accreting supermassive black hole produced a significant outflow that vacated a sufficiently large cone on scales of 10-100 pc---fortuitously aligned in our direction---that the quasar has become clearly visible at rest-frame UV wavelengths, while the host galaxy is not. 
Significant outflows have indeed been observed from high-redshift quasars \citep[e.g.][]{Alexander2010,Nesvadba2011,maiolino_2012}, and are expected to be able to carve a window for observing the quasar through otherwise high-density gas  \citep{Ni2019}. Thus, it seems likely that such outflows are present in these systems. These objects are therefore high priority targets for JWST, which will add 1.6--29 $\mu$m wavelength imaging coverage to our 1.2--1.6 $\mu$m HST images, and so is expected to much better constrain the dust extinction and geometry that UV images alone cannot capture.

\subsection{Quasar Selection Bias}
\label{SelectionBias}
Five \zsim{6} quasars for this HST program were selected as those UV-faint quasars with confirmed sub-mm detections, and thus the greatest rest-frame $L_{\textrm{FIR}}/L_{\textrm{UV}}$ ratios. 
This selects for host systems with the greatest non-AGN contribution to the FIR flux, with inferred ultraluminous infrared galaxy (ULIRG)-class FIR luminosities ($>10^{12}~\LSun{}$) and implied star formation rates of $\approx500M_\odot \textrm{yr}^{-1}$ \citep{Wang2011}. 
Locally, ULIRGs are gas-rich with high inferred star formation rates, and most are undergoing major mergers or at least strong interactions \citep[e.g.][]{Howell2010,Kim2013}. This suggests that this sample of $z\simeq6$ quasars are a distinct quasar sub-population, which may be biased towards quasars with nearby interacting galaxies. 
This potential selection bias may mean that while our six quasars are in environments typical of luminous $z\simeq6$ galaxies, the overall $z\simeq6$ quasar population may reside in somewhat under-dense environments.
However, note that \citet{Trakhtenbrot2018} observed three FIR-bright and three FIR-faint quasars with ALMA, finding spectroscopically confirmed sub-mm companion galaxies interacting with three quasars---one FIR-bright and two FIR-faint. We also find a potential companion around SDSS J0005-0006, which is not detected in the FIR \citep{Wang2008,Jiang2013,leipski_2014}. Hence, companion galaxies are not necessarily a feature of only FIR-bright quasars. 

This selection bias is also likely to affect the measured black hole--stellar mass relation, as these quasars are not necessarily representative of the overall $z\simeq6$ quasar population. 
For example, the most highly star-forming quasar host galaxies in BlueTides show a significantly steeper black hole--stellar mass relation, with such quasars lying on the main relation for the most massive black holes, and below the relation for lower-mass black holes. This result suggests that our ULIRG-type hosts, which are also selected to be UV-faint and thus potentially have lower mass black holes, may lie below the black hole--stellar mass relation of the full quasar population. 

The bias to selecting ULIRG-class host galaxies may also affect our stellar mass limits, as these galaxies generally lie significantly above the SFR--stellar mass main sequence. Their extreme star-formation rates may indicate the presence of large amounts of dust extinction, as discussed in Section \ref{sec:Dust}, which could further bias our measurements.

\subsection{The Prevalence of $z\simeq6$ Quasars with Companions}
Companion galaxies have been discovered around a range of high-redshift quasars, with the majority seen only in observations at sub-mm wavelengths \citep[e.g.][]{Wagg2012,Decarli2017}. 
For example, \citet{Trakhtenbrot2017,Trakhtenbrot2018} found companions physically associated with three of six $z\simeq4.8$ quasars observed with ALMA, at separations of 14-45 kpc. Those companions have dynamical masses $M_{\textrm{dyn}}=(2.1-10.7) \times 10^{10} M_\ast$, compared with the quasar hosts which have $M_{\textrm{dyn}}=(3.7-7.4) \times 10^{10} M_\ast$, indicative of major galaxy interactions.
These companion galaxies are not detected in Spitzer data, and so \citet{Trakhtenbrot2018} conclude that there must be significant dust-obscuration.
This result may explain the lack of companions observed in rest-frame UV observations \citep[e.g.][]{willott_2005}; this is supported by simulations \citep{Marshall2019}.

Our potential companion galaxies are detected in the rest-frame UV, suggesting that these companions may have less dust attenuation than those observed in the sub-mm. Other studies have also observed companions in the rest-frame UV; for example, \citet{McGreer2014} discovered a companion galaxy around both a $z = 4.9$ and a $z = 6.25$ quasar, at 5 and 12~kpc projected separations. While the companion of the $z = 4.9$ quasar is spectroscopically confirmed, the $z = 6.25$ companion is presumed to be at that redshift based on imaging in two HST filters, as in our study.
While identifying these two companions, \citet{McGreer2014} reported that bright companions around high-redshift quasars are uncommon, with an incidence of $\lesssim2/29$ for $\gtrsim 5L^\ast$ galaxies and $\lesssim1/6$ for $2\lesssim L \lesssim 5L^\ast$ galaxies.

Finding quasar companion galaxies is consistent with the scenario that the growth of high-redshift quasars is triggered by galaxy mergers. 
While simulations predict that galaxy mergers can fuel quasar activity, it is unclear if these are the dominant cause of high-redshift black hole growth. For example, using the EAGLE simulation, \citet{McAlpine2018} reported that at $z=0$, $\sim60\%$ of black holes undergoing a rapid growth phase do so within $\pm0.5$ dynamical times of a galaxy-galaxy merger, and \citet{McAlpine2020} found an over-abundance of AGN within merging systems relative to control samples of inactive or isolated galaxies. However, while galaxies experiencing mergers have two to three times higher accretion rates than isolated galaxies, the majority of black hole mass growth does not occur during the merger periods \citep{McAlpine2020}.
Thus, if the potential companions are confirmed to be associated with our $z\simeq6$ quasars, this result may be due to our biased sample of FIR-luminous quasars (see discussion in Section \ref{SelectionBias}) and not necessarily indicative that nearby companions are common around high-redshift quasars.

\section{Summary}\label{conclusions}

We use Hubble Space Telescope imaging of five far infrared-luminous $z\simeq6$
quasars, and the hot-dust free quasar SDSS J0005-0006, to search for rest-frame UV emission from their host galaxies.
Using the Markov Chain Monte Carlo estimator \soft{psfMC}, we perform 2D surface brightness modeling for each quasar to model and subtract the quasar point source in order to detect possible underlying host emission. 

Only upper limits were found for the quasar host galaxies, of $m_J>22.7$ and $m_H>22.4$ mag.
These limits are beginning to probe magnitudes expected for high-redshift quasar hosts from the BlueTides simulation, which suggests that the increased resolution and near--mid-infrared spectroscopic capability of the James Webb Space Telescope (JWST) should detect host emission in the rest-frame UV/optical for the first time \citep[see also the BlueTides predictions of][]{Marshall2019}. We also expect that these host galaxies could be quite dusty, with $\langle A_{\mathrm{UV}}\rangle\simeq2.6$ mag (see Figure \ref{fig:BT_comparison}), and thus probing their mid-infrared emission with JWST will be invaluable.

Converting these magnitude limits to stellar mass limits suggests that five of the six quasars could be consistent with the local black hole--stellar mass relation of \citet{kormendy_2013}, and with existing sub-mm observations of $z\simeq6$ quasar hosts \citep{Willott2017,Izumi2018,Izumi2019,Pensabene2020}.
SDSS-J0203+0012 has a stellar mass of $\log(M_\ast/M_\odot)<11.28$ and a large black hole mass of $\log(M_{\textrm{BH}}/M_\odot)=10.72$, which places it above the local relation. However, its black hole mass is likely inaccurate.

We detect up to nine potential $z\simeq6$ companion galaxies surrounding five of the six quasars, with magnitudes and UV spectral slopes consistent with luminous $z\simeq6$ star-forming galaxies. These galaxies lie within $1\farcs4$--$3\farcs2$ of the quasars, or at a projected distance of 8.4--19.4 kpc (if at the same redshift). 
If their true distance is of order their projected distance, these companions could be interacting with the quasar host galaxies, potentially enhancing their quasar activity  \citep{McAlpine2020,Patton2020}.
Finding nine potential $z\simeq6$ companion galaxies is consistent with expectations for large-scale clustering around high-redshift quasars \citep{Garcia2017} and galaxies \citep[][see Figure  \ref{fig:overdensity}]{Qiu2018}.
Hence, we find that our quasars are in environments typical of luminous $z\simeq6$ galaxies.

The existing data cannot rule out the probability that some of these potential companions are foreground interlopers.
Future observations will focus on better constraining the spectral energy distributions of the companions, including deep $r$-band imaging to identify low-redshift interlopers, and adaptive optics-corrected K-band imaging to better constrain the rest-frame UV SED. The launch of the JWST will allow spectroscopic measurements of the redshifts of these potential companion galaxies, determining whether they are indeed physically associated with the quasars, and perhaps being high-redshift major mergers in progress.

\acknowledgements
{We thank the anonymous referee for their valuable feedback, which helped to improve the quality of this paper. 
We thank Tony Roman, Tricia Royle and the Space Telescope Science Institute staff for their continued
excellent help in scheduling our non-standard HST observations---they went
beyond the call of duty to make the best possible HST data happen. We
acknowledge support provided by NASA through grants GO-12332.*A, GO-12974.*A, and
GO-12613.*A from the Space Telescope Science Institute, which is operated by the
Association of Universities for Research in Astronomy, Inc., under NASA
contract NAS 5-26555. This work was supported by NASA JWST Interdisciplinary
Scientist grants NAG5-12460, NNX14AN10G, and 80NSSC18K0200 to RAW from GSFC. This research was supported by the Australian Research Council Centre of Excellence for All Sky Astrophysics in 3 Dimensions (ASTRO 3D), through project number CE170100013. MAM acknowledges the support of an Australian Government Research Training Program (RTP) Scholarship.
MM and KJ acknowledge support through the Extraterrestrische Verbundforschung program of
the German Space Agency, DLR, grant number 50 OR 1203. 

This work was partially performed on the OzSTAR national facility at Swinburne University of Technology. OzSTAR is funded by Swinburne University of Technology and the National Collaborative Research Infrastructure Strategy (NCRIS). 
This research made use of Python packages NumPy \citep{Numpy2011}, Matplotlib \citep{Matplotlib2007}, AstroPy \citep{Astropy2013}, Pandas \citep{reback2020pandas}, SciPy \citep{2020SciPy-NMeth}, emcee \citep{emcee2013}, and corner \citep{corner}.
}

\bibliography{manuscript.bib}

\appendix
\setlength{\parindent}{0pt}
\section{Magnitude Calculation}
\label{Appendix}
To convert the surface brightness in the annulus 0$\farcs$42--0$\farcs$60 to a magnitude, we consider the host galaxy to have a S\'ersic profile and to be azimuthally symmetric.
The total flux contained within radius $R$ is 
\begin{equation}
F(R) = \int_0^R 2\pi r f(r) dr
\end{equation}
where $f(r)$ is the flux per unit physical area at radius $r$. 
For a S\'ersic profile with S\'ersic index $n$ and effective radius $R_e$, 
\begin{equation}
f(r)=f(R_e) \exp \left( -b_n \left((r/R_e)^{1/n} -1\right)\right)
\end{equation}
where $b_n$ is defined to satisfy $\Gamma(2n)=2\gamma(2n,b_n)$, where $\Gamma$ and $\gamma$ are the complete and incomplete gamma functions, $\Gamma(a)=\int_0^\infty t^{a-1} e^{-t} dt$ and $\gamma(a,x)=\int_0^x t^{a-1} e^{-t} dt$.
Thus, $F(R)$ can be expressed as
\begin{equation}
F(R) = 2 \pi R_e^2 f(R_e)~ n e^{b_n} b_n^{-2n} ~ \gamma(2n,b_n (R/R_e)^{1/n}).
\end{equation}
Given $\Delta F=F(0\farcs60)-F(0\farcs42)$ is the flux measured in the annulus, $f(R_e)$ can be calculated from:
\begin{equation}
f(R_e) = \Delta F \left[2 \pi R_e^2  n e^{b_n} b_n^{-2n}~ \left[\gamma(2n,x_1)-\gamma(2n,x_2)\right]\right]^{-1}
\end{equation}
where $x_1=b_n (0.60/R_e \textrm{[arcsec]})^{1/n}$ and  $x_2=b_n (0.42/R_e \textrm{[arcsec]})^{1/n}$.
We thus calculate the flux of the host galaxy as $F(0\farcs60)$.

\begin{figure}
\centering
\includegraphics[scale=1]{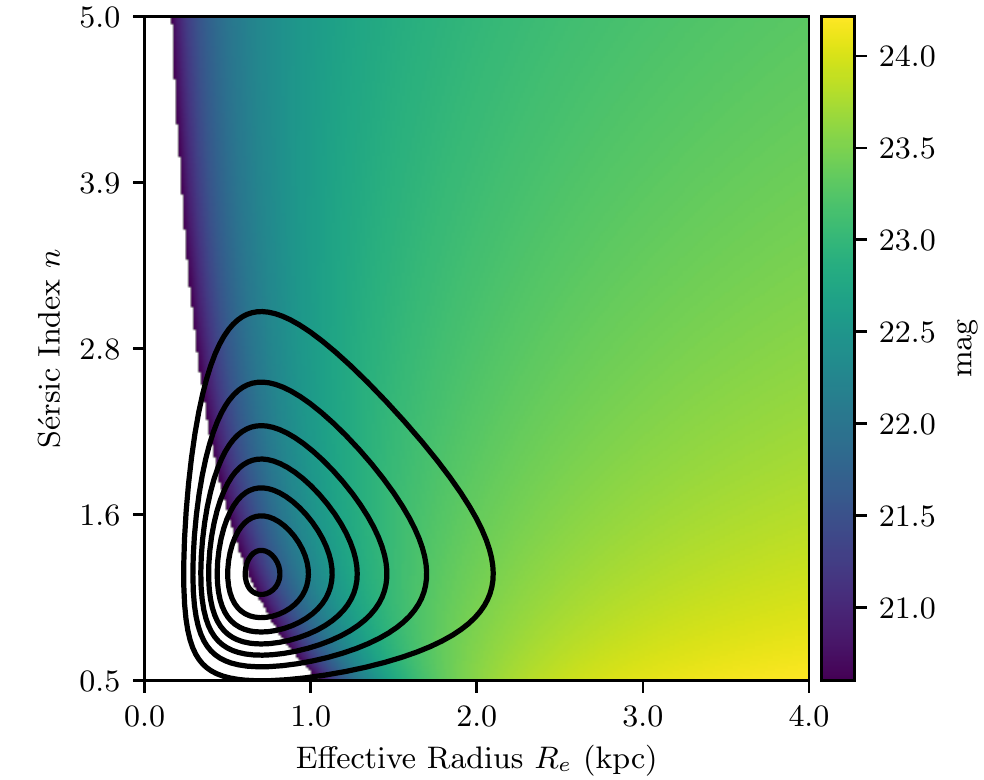}
\caption{The J-band magnitude of the host galaxy of NDWFS-J1425+3254, as an example, assuming a S\'ersic profile with S\'ersic index $n$ and effective radius $R_e$ that is constrained to have the measured flux (a $2\sigma$ noise limit) in the annulus 0$\farcs$42--0$\farcs$60. We show a range of S\'ersic profiles, with $n \in (0.5,5)$ and $R_e \in (0,4)$ kpc. 
The white regions show galaxies with magnitudes brighter than the quasar itself (here $m_J<20.6$), which we exclude.
The black contours show the 2D probability distribution functions for $n$ and $R_e$, guided by the observations of $z\simeq6$ bright (1--10$L^\ast_{z=3}$), massive galaxies by \citet{Shibuya2015}, assuming the parameters are independent. Using a Monte Carlo technique to sample magnitudes from this distribution gives a magnitude probability distribution function, from which we choose the most likely value as the corresponding magnitude limit, in this case $m_J>22.8$ mag. 
}
\label{Appendix:Sersic}
\end{figure}

We calculate this flux for a range of S\'ersic profiles, with $n \in (0.5,5)$ and $R_e \in (0,4)$ kpc, ignoring galaxies which have magnitudes brighter than the observed magnitude of the quasar.  We show an example for the J-band magnitude of NDWFS-J1425+3254 in Figure \ref{Appendix:Sersic}.

Guided by the observations of $z\simeq6$ bright (1--10$L^\ast_{z=3}$), massive galaxies by \citet{Shibuya2015}, we assume probability distribution functions for $n$ and $R_e$, with the combined 2D probability distribution function shown in Figure \ref{Appendix:Sersic}. We use a Monte Carlo technique to sample from this distribution, and determine the resulting probability distribution function for host galaxy magnitude. We choose the most-likely value from this distribution as our magnitude limit. Note that this is not the magnitude of the most likely $n$-$R_e$ combination, as many less-likely $n$-$R_e$ combinations produce similar magnitudes and make those more likely.

\end{document}